\documentclass{article}
\usepackage{ijcai24}
\usepackage{paralist}

\usepackage{times}
\usepackage{soul}
\usepackage{url}

\usepackage[utf8]{inputenc}
\usepackage[small]{caption}
\usepackage{graphicx}
\usepackage{amsmath, amsfonts}
\usepackage{amsthm}
\usepackage{booktabs}
\usepackage{algorithm}
\usepackage{algorithmic}
\usepackage[switch]{lineno}

\usepackage{natbib}

\usepackage{multirow}
\usepackage{graphicx}
\usepackage{array}
\usepackage{makecell}
\usepackage{longtable}
\usepackage{pdflscape}
\usepackage{rotating}

\usepackage{color, xcolor}

\newcommand{\bap}[1]{\textcolor{blue}{@Aditya:~#1@}}

\let\cite\citep

\usepackage[hidelinks]{hyperref}
\hypersetup{
    colorlinks=true,
    citecolor=black
          }

\title{A Comprehensive Survey on Graph Reduction: Sparsification, Coarsening, and Condensation}

\author{
Mohammad Hashemi\footnote{Equal contribution.}$^1$
\and
Shengbo Gong$^*$$^1$\and
Juntong Ni$^1$ \and \\
Wenqi Fan$^{2}$\and 
B. Aditya Prakash$^{3}$ \and
Wei Jin$^{1}$
\affiliations
$^1$Emory University, \\
$^2$The Hong Kong Polytechnic University, \\
$^3$Georgia Institute of Technology\\
\emails
\{mohammad.hashemi, shengbo.gong, juntong.ni, wei.jin\}@emory.edu,
wenqi.fan@polyu.edu.hk,
badityap@cc.gatech.edu
}

\begin{document}

\maketitle

\begin{abstract}
Many real-world datasets can be naturally represented as graphs, spanning a wide range of domains. However, the increasing complexity and size of graph datasets present significant challenges for analysis and computation. In response, graph reduction, or graph summarization, has gained prominence for simplifying large graphs while preserving essential properties. In this survey, we aim to provide a comprehensive understanding of graph reduction methods, including graph sparsification, graph coarsening, and graph condensation. Specifically, we establish a unified definition for these methods and introduce a hierarchical taxonomy to categorize the challenges they address. Our survey then systematically reviews the technical details of these methods and emphasizes their practical applications across diverse scenarios. Furthermore, we outline critical research directions to ensure the continued effectiveness of graph reduction techniques, as well as provide a comprehensive paper list at \url{https://github.com/Emory-Melody/awesome-graph-reduction}. We hope this survey will bridge literature gaps and propel the advancement of this promising field.
\end{abstract}

\section{Introduction} 
Graph-structured data has become ubiquitous in various domains, ranging from social networks and biological systems to recommendation systems and knowledge graphs~\citep{fan2019socialreco,wu2022reco,wu2018leveraging,shi2017KG,wang2021leverage}. The inherent relational structure of graph data makes it a powerful representation for modeling complex interactions and dependencies. Moreover, with the rise of graph machine learning techniques, especially graph neural networks (GNNs)~\citep{kipf2016semi,wu2020comprehensive,ma2021deep}, the utilization of graph datasets has seen unprecedented growth, leading to advancements in fields such as chemistry~\cite{reiser2022graph}, biology~\cite{gligorijevic2021structure}, and epidemiology~\citep{liu2024review}.


\begin{figure}[t] 
  \centering
  \includegraphics[width=0.5\textwidth]{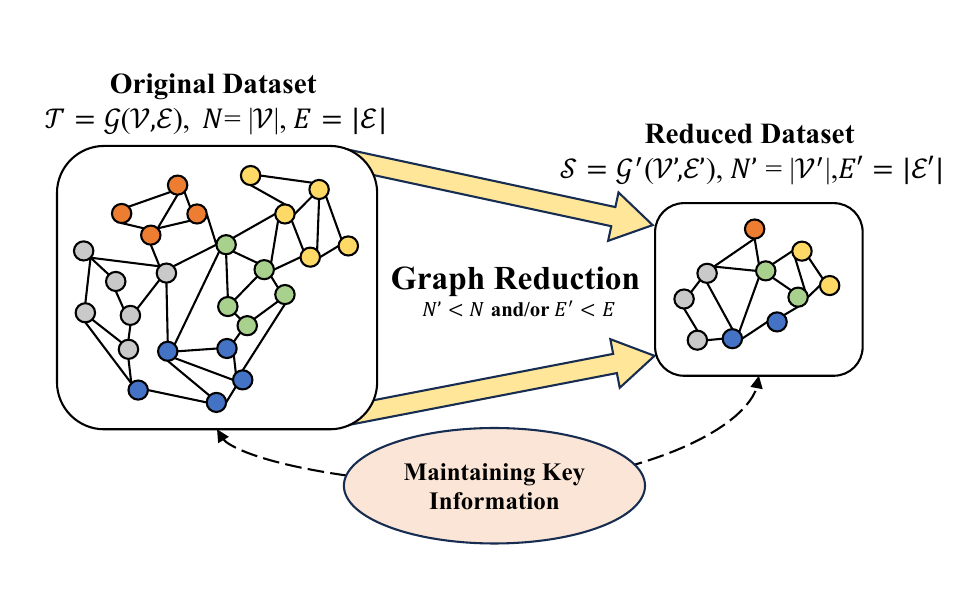} 
  \vskip -2em
  \caption{A general framework of graph reduction. Graph reduction aims to find a reduced (smaller) graph dataset that can preserve certain information of the original graph dataset.}
  \label{fig:intro}
  \vskip -1em
\end{figure}

Recent years have witnessed an exponential increase in the size and complexity of graph datasets. Large-scale networks, such as social graphs and citation networks~\cite{hu2021ogb}, challenge the scalability and efficiency of existing algorithms and demand innovative solutions for efficient model training. Despite recent efforts to design GNNs that can scale with large graphs~\citep{jia2020redundancyfree,zeng2021decoupling,song2023xgcn,liu2021exact}, an alternative approach focuses on reducing the size of the graph dataset, including the number of graphs, nodes, and edges, which we term as \textbf{graph reduction}\footnote{It is also known as graph summarization, simplification or degeneracy in some literature. We choose to consistently use ``graph reduction'' throughout this survey for clarity and uniformity.}~\cite{jin2021gcond,huang2021scaling}. 
In this paper, we define graph reduction as \textit{the process of finding a graph dataset of smaller size while preserving its key information}.
Specifically, this definition requires an algorithm that takes the original graph dataset as input and produces a smaller one. As shown in Figure~\ref{fig:intro}, graph reduction aims to extract the essential information from the massive graph dataset by maintaining its structural and semantic characteristics. 
In addition to accelerating graph algorithms, graph reduction offers a range of advantages. \textbf{First}, reduced graphs demonstrate compatibility with various downstream models architectures~\cite{jin2021gcond}.
\textbf{Second}, graph reduction may contribute to privacy preservation since it alters the original structure or node attributes, making them challenging to recover~\cite{dong2022privacyforfree}.
\textbf{Third}, the reduced graph is notably smaller and more comprehensible for humans compared to its larger counterpart, which aids in graph visualization~\cite{imre2020spectrumvisualize}. 

\begin{figure*}[t] 
  \centering
  \includegraphics[width=1\textwidth]{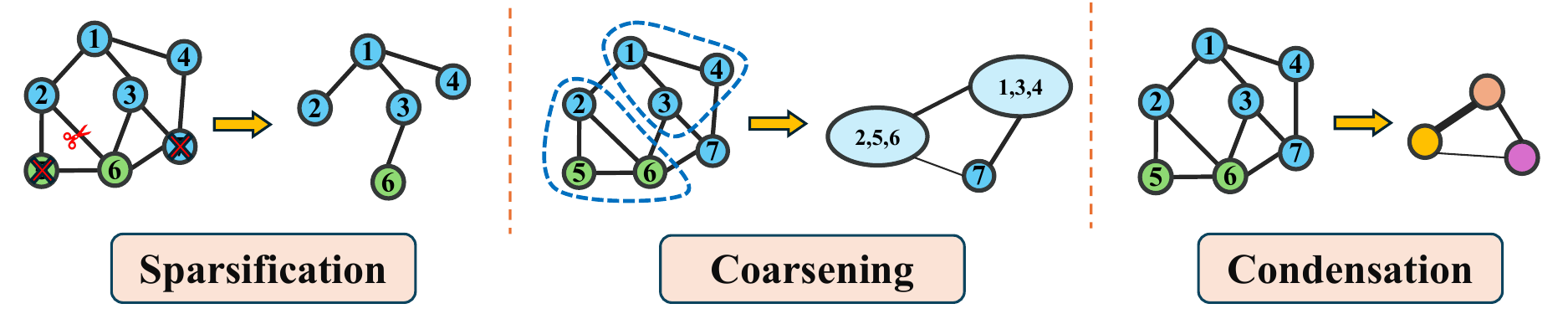} 
  \caption{Illustration of key differences among three strategies of graph reduction: Graph sparsification selects significant nodes and edges while discarding others, graph coarsening groups and aggregates similar nodes and edges to construct a smaller graph, and graph condensation learns a synthetic graph from scratch.}
  \label{fig:illustration}
\end{figure*}

Given the importance of graph reduction, numerous algorithms have been developed, falling into three distinct strategies: \textbf{graph sparsification}~\citep{althofer1993spannerprove,batson2009twice}, \textbf{graph coarsening}~\citep{loukas2018spectrally,dorfler2012kron}, and the more recent \textbf{graph condensation}
~\citep{jin2021gcond,jin2022doscondensing,xu2023kernel,liu2022graph}.  Graph sparsification revolves around the approximation of a graph by retaining only a subset of its edges and vital nodes. In contrast, graph coarsening does not eliminate any nodes but instead groups and amalgamates nodes into super nodes, with original inter-group edges being aggregated into super edges using a specified aggregation algorithm. Differing from the aforementioned two strategies, graph condensation has been recently introduced as a method to condense a graph by synthesizing a smaller graph while preserving the performance of GNNs. Despite the proliferation of these methods, they have often been studied in isolation, leaving the connections and distinctions between them somewhat obscured. Therefore, it is both necessary and timely to offer a systematic overview of these existing algorithms in order to enhance our understanding of graph reduction techniques.

\vskip 0.3em
\noindent\textbf{Contributions.}  In this work, we aim to present a comprehensive and up-to-date survey focusing on graph reduction techniques and their diverse applications in tackling graph-related challenges. We aspire for this survey to serve as a valuable resource for both novice researchers and practitioners interested in exploring this field, while also catalyzing future research endeavors. Our contributions can be summarized as follows:  
\begin{compactenum}[(a)]
    \item We offer the first comprehensive review of graph reduction methods, encompassing graph sparsification, graph coarsening, and graph condensation.
    \item We develop a unified perspective for existing graph reduction methods, differentiating them based on their characteristics in Section~\ref{sec:taxonomy}, and provide a detailed review of representative algorithms in Section~\ref{sec:method}.
    \item We discuss practical applications of graph reduction methods in Section~\ref{sec:app}, shedding light on real-world scenarios where these techniques prove valuable.
    \item In Section~\ref{sec:future}, we identify key challenges and promising future research directions, guiding the continued advancement of graph reduction techniques.
\end{compactenum}

\vskip 0.3em
\noindent\textbf{Connection to Existing Surveys}. 
In contrast to previous reviews on graph reduction~\citep{liu2018summarization,interdonato2020multilayer,shabani2023surveyGNN,chen2022graphscsurvey}, our study offers a comprehensive overview of the emerging field of graph condensation and presents a unified framework that bridges graph condensation with conventional graph reduction techniques. Compared with the latest survey on graph learning acceleration~\cite{zhang2023surveyacceleration} which also categorizes graph reduction algorithms into those as mentioned above three, our work uses more general definitions and discusses more related methods on each category in a wider scope.
Additionally, our survey explores synergies between graph reduction and GNNs, an aspect rarely covered in existing surveys. 
Also, some data-centric graph learning surveys~\cite{zha2023datacentric,zheng2023towardsdatacentric}include discussions on graph reduction but we offer a more detailed and thorough examination of reduction techniques. Furthermore, our work shares connections with recent surveys on dataset distillation~\citep{geng2023surveydatadistillation,sachdeva2023DDasurvey}, while they majorly focus on condensation methods applied to image data.

\begin{table}[h]
\caption{Notations used in this paper.} 
\label{tab:notations}
\centering
\begin{tabular}{ll}
\toprule
\textbf{Notation} & \textbf{Description} \\
\hline
$G$ & A graph \\
$G'$ & A reduced graph \\
$\mathcal{V}$ & Set of graph nodes \\
$\mathcal{E}$ & Set of graph edges \\
$\bf X$ & Node feature matrix \\
$\bf A$ & Adjacency matrix \\
$\bf Y$ & One-hot label matrix \\
$\bf L$ & Graph Laplacian matrix\\
$\bf C$ & Node assignment/mapping matrix\\
$\mathcal{T}$ & Original graph dataset \\
$\mathcal{S}$ & Synthetic graph dataset \\
$N$ & Number of nodes \\
\bottomrule
\end{tabular} 
\end{table}

\section{Taxonomy of Graph Reduction} \label{sec:taxonomy}



Before we formally introduce the definition of graph reduction, we first introduce the notations used in this paper. Given the node set $\mathcal{V}$ and edge set $\mathcal{E}$, we denote a graph as $G=(\mathcal{V},\mathcal{E})$. In attributed graphs, nodes are associated with features, and thus can be represented as $G=({\bf A}, {\bf X})$, where
${\bf X}=[{\bf x}_1,{\bf x}_2,...,{\bf x}_N]$ denotes the node attributes and ${\bf A}$ denotes the adjacency matrix. The graph Laplacian matrix is $\bf L=\bf D-\bf A$, where $\bf D$ is a diagonal degree matrix with ${\bf D}_{ii}=\sum_j {\bf A}_{ij}$. We use $N=|\mathcal{V}|$ and $E=|\mathcal{E}|$ to denote the number of nodes and edges, respectively. We summarize the major notations in Table~\ref{tab:notations}.

\paragraph{A Unified Framework of Graph Reduction} 
Given a graph $G=(\mathcal{V},\mathcal{E})$,  graph reduction outputs a graph ${G}'=(\mathcal{V}',\mathcal{E}')$ which contains $N'$ nodes and $E'$ edges, subject to $N'< N$, or $E'< E$ edges. The reduced graph $G'$  preserves the desired information of the original graph $G$. This process can be understood as finding a graph $G'$ that minimizes a loss function $\mathcal{L}$ explicitly or implicitly, which measures the difference between $G$ and $G'$ in terms of certain information:
\begin{equation}
{G'}=\underset{{G'}}{\arg \min } \ \mathcal{L}({G}, {G'}). \label{eq:graphReduction}
\end{equation}

\textit{Remark 1.} Note that the desired outcome of graph reduction should be represented as a graph, leading to the exclusion of graph representation learning methods~\cite{wu2020comprehensive} in the context of graph reduction.  Furthermore, we restrict this definition to graphs that converge to an optimal state through the algorithmic process. This distinction sets it apart from data augmentation techniques~\cite{rong2019dropedge,zhao2022graph}, where the augmented graphs vary with each epoch.

\textit{Remark 2.} While the majority of graph reduction methods focus on decreasing the number of nodes or edges within a graph, there are also studies~\cite{jin2022doscondensing,xu2023kernel} that reduce the number of distinct graphs, particularly in applications such as graph classification. In this survey, unless explicitly specified otherwise, our primary focus remains on the former approach, as it is more commonly employed.

On top of that, we can categorize existing graph reduction techniques into the following three groups, based on how they produce the reduced graphs:
\paragraph{Definition of Graph Sparsification} 
Given a graph $G = ({\bf A}, {\bf X})$, graph sparsification selects existing nodes or edges from the graph $G$ and outputs $G'=({\bf A}',{\bf X}')$. In other words, the elements in ${\bf A}'$ or ${\bf X}'$ are the subset of those in ${\bf A}$ or ${\bf X}$.


\paragraph{Definition of Graph Coarsening}
Given a graph $G=({\bf A}, {\bf X})$, graph coarsening outputs $G'=({\bf A}',{\bf X}')$ containing $N'$ super-nodes and $E'$ super-edges, where $N'<N$. It requires finding a surjective mapping from the original graph to a coarse graph, which can be formulated by a one-hot matrix ${\bf C}\in \{0,1\}^{N \times N'}$ that assigns nodes to super-nodes.
We further define the reverse assignment matrix ${\bf P}=\operatorname{rowNormalize}({\bf C}^\top)$. Then the coarse graph is usually constructed by ${\bf A}'={\bf C}^\top{\bf A}{\bf C}, {\bf X}'={\bf P}{\bf X}$ with the Laplacian matrix being ${\bf L}'={\bf C}^\top{\bf L}{\bf C}$.

\paragraph{Definition of Graph Condensation}\label{subsec:graphCondensation}


Given a graph $\mathcal{T}=(\mathbf{A}, \mathbf{X}, \mathbf{Y})$, with ${\bf Y}$ being the node labels,
graph condensation aims to learn a small-synthetic graph dataset $\mathcal{S}=(\mathbf{A'}, \mathbf{X'}, \mathbf{Y'})$, where $\mathcal{S}$ contains learnable parameters, and $N'<N$, such that a GNN trained on $\mathcal{S}$ obtains a comparable performance to the one trained on $\mathcal{T}$.



\begin{table*}[t]
\centering
\small 
\caption{General qualitative comparison of graph reduction methods. ``Scalability'': the ability to scale up to large graphs, ``Interpretability'': the existence of correspondence between nodes in the original and reduced graphs, ``Label Utilization``: the reliance on label information.}
\label{tab:comparison}
\renewcommand{\arraystretch}{1}
\centering
\small
\resizebox{\textwidth}{!}{
\begin{tabular}{lllllll} 
\toprule
\textbf{Strategy}  & \textbf{Interpretability} & \textbf{Label Utilization} & \textbf{Objective} & \textbf{Output} & \textbf{Remark}\\ \hline
Sparsification & \checkmark & $\times$ & Property/Performance Preservation & Subgraph & Select subgraphs with maximal information\\
Coarsening  & \checkmark  & $\times$ & Property/Performance Preservation & Supergraph & Merge close nodes\\
Condensation  & $\times$ & \checkmark & Performance Preservation & Synthetic graph & Generates small graphs with high utility\\
\bottomrule
\end{tabular}}
\end{table*}

\vskip 0.3em
\noindent\textbf{Distinctions.}  
The above three strategies share a common goal: to obtain a small graph that can preserve key information and benefit downstream processes. However, they differ in three key aspects. \textbf{First}, graph condensation synthesizes fake graph elements, while sparsification selects existing ones and coarsening aggregates them. The latter two strategies enjoy certain interpretability in the reduction process, as the reduced graph can be easily understood and related back to the original graph.
\textbf{Second}, these strategies have distinct objectives. Graph condensation aims to maintain the performance of GNN models in downstream tasks, while the other two often target at preserving graph properties.
\textbf{Third}, graph condensation relies on labels, whereas the other two generally do not.

\begin{figure*}[t] 
  \centering
  \includegraphics[width=1\textwidth]{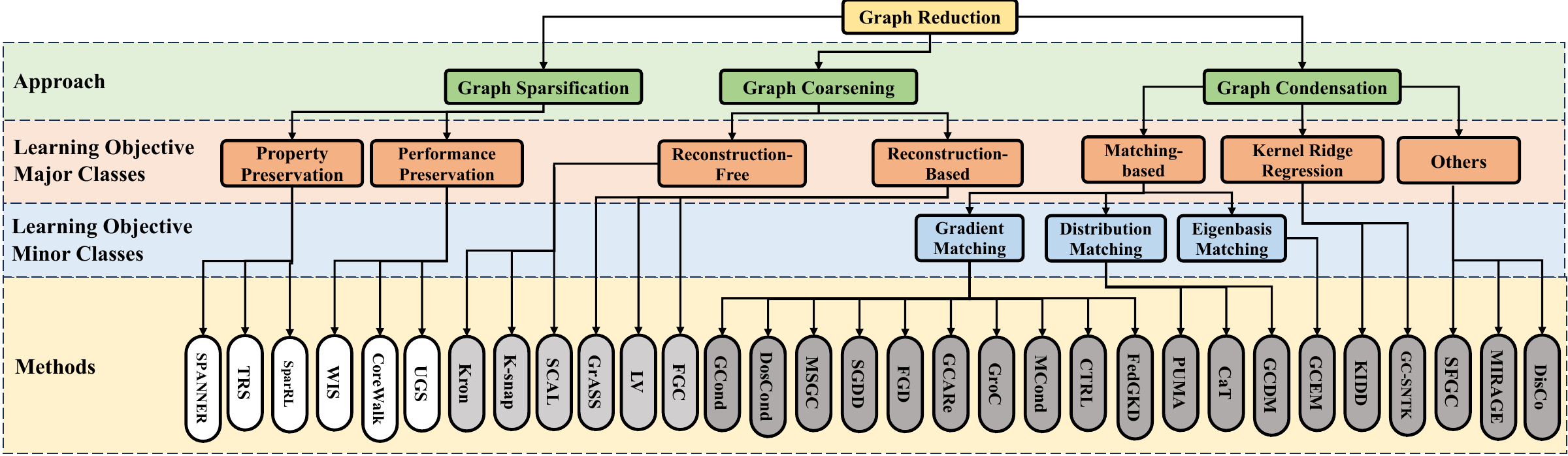} 
  \caption{Taxonomy of existing graph reduction methods.}
  \label{fig:taxonomy}
\end{figure*}

In Figure~\ref{fig:taxonomy}, we present a detailed taxonomy of existing graph reduction methods within the aforementioned categories, and we will elaborate in the following section. Additionally, Table \ref{tab:comparison} provides a qualitative comparison of the three graph reduction strategies mentioned earlier.

\section{Methodology} \label{sec:method}
In this section, we introduce the representative algorithms for the aforementioned three strategies of graph reduction. For each strategy, we categorize methods by their learning objectives and summarize popular approaches in Table~\ref{tab:taxonomy}.
\subsection{Graph Sparsification}\label{subsec:graphSparsification}
Graph sparsification, as an intuitive method for graph reduction, involves selecting essential edges or nodes based on specific criteria and then constructing a smaller graph from these selected elements. Conventional approaches typically focus on preserving specific graph properties, such as spectrum and centrality. With the increasing popularity of GNNs, there is a growing array of methods aimed at maintaining the quality of node representations. Consequently, we categorize existing techniques into two groups based on their preservation goals: those focused on preserving graph properties and those dedicated to maintaining model performance.




\subsubsection{Preserving Graph Properties}
In traditional graph sparsification, essential graph properties include pairwise distances, cuts, and spectrum~\cite{batson2013sparsifier}. Sparsification methods iteratively sample the subgraphs that achieve the minimal loss $\mathcal{L}(G', G)$ in a greedy manner, which measures the approximation to the original graph w.r.t. one of the above graph properties.
A reduced graph is called \textit{spanner} if it maintains pairwise distances, and \textit{sparsifier} if it preserves cut or spectrum~\cite{batson2013sparsifier}.
To evaluate these algorithms, one common way is to establish the loss bound for their output graph $G'$: If $G'$ is proved to satisfy $\mathcal{L}(G', G)\le \epsilon,\epsilon\in(0,1)$, it is called $\epsilon$-spanner/sparsifier. Specifically, $\mathcal{L}(G',G)$ is expressed as $|D(G', G)-1|$ with $D(\cdot,\cdot)$ defined as follows:
\begin{equation}
    D(G', G)=\begin{cases}\frac{\operatorname{SP}(G')}{\operatorname{SP}(G)}&\text{for spanner}, \\
\frac{\bf{x}^\top\bf{L}'\bf{x}}{\bf{x}^\top\bf{L}\bf{x}}&\text{for sparsifier}, \end{cases}
\end{equation}
where $\operatorname{SP}(G)$ denotes the sum of the shortest path length for all node pairs in $G$, and ${\bf x} \in  \mathbb{R}^N$ is an arbitrary vector.

\textit{Spanners.} 
\citet{althofer1993spannerprove} first develop an algorithm named SPANNER to obtain spanners in graphs. It starts with an empty graph defined on the original node set and adds edges from the original graph only if their weight is smaller than the current distance between their connected nodes in the reduced graph. They also prove that every weighted graph has a $2t$-spanner with $O(N^{1+1/t})$ edges. 
~\citet{baswana2003simplespanner} tighten this upper bound to $(2t-1)$ with a linear time algorithm that merely explores the edges in the neighborhood
of a node. Furthermore, by defining a reinforcement learning process and adjusting reward functions to the preservation of pairwise distance, SparRL~\cite{wickman2022sparsebyRL} outperforms all conventional graph sparsification approaches. 

\textit{Sparsifiers.}
One representative sparsifier is called \textit{Twice Ramanujan Sparsifier} (TRS)~\citep{batson2009twice}, which prove that for every $\epsilon\in (0, 1)$ and every undirected graph $G$, there exists a weighted graph $G'$ with at most $(N-1)/2$ edges such that $G'$ is the $(1+\epsilon)$-sparsifier of $G$ with high probability. 
This approach presents an algorithm for deriving $G'$ by decomposing the graph into subgraphs with high \textit{conductance}, calculating pairwise effective resistance (ER)~\citep{spielman2008effectiveresistance}, and sampling edge based on normalized ER as probabilities. Then the edges in the reduced graph are reweighted as the probabilities.
Furthermore, ~\citet{lee2018twicelinear} presents an almost-linear time algorithm for constructing such a sparsifier.  
Previous studies typically necessitate alterations to edge weights as part of the reweighting process. \citet{anderson2014unweighted} address the sparsifier problem conditioned by keeping the original edge weights, employing a method of unweighted column selection. 
Since most theories are constructed upon the undirected graph, \citet{chung2005laplacianssysmentricdirected} first extends them into directed graphs by first symmetrizing the graph Laplacians. 
Different from the above methods that only cut edges, \citet{feng2016spectral} first finds an extremely sparse subgraph -- low-stretch spanning tree, and recovers small portions of off-tree edges to further approximate the spectrum. 
GSGAN~\cite{wu2020graphgan} designs a reward
function to guide the graph generator in a generative adversarial network creating random walks. These random walks are finally combined to form a smaller edge set that is effective for community detection.


\subsubsection{Preserving Performance}
With the emergence of GNNs, a new goal of graph sparsification has arisen: maintaining the prediction performance of GNNs trained on the sparsified graph. In this context, the sparsification process selects the top-$k$ nodes or edges based on various scoring methods, such as ER~\citep{spielman2008effectiveresistance}, PageRank~\citep{langville2004deeperpagerank}, KCenter~\citep{sener2018activecoreset} and explanations from a trained GNN~\cite{ying2019gnnexplainer}.

Many methods employ model-free heuristics as the scoring strategy, which calculate the score with metrics derived from the graph structure.  
For example, \citet{salha2019kcoredegeneracy} use $k$-core decomposition to find interconnected subgraphs with different density index $k$. By treating subgraphs corresponding to high values of $k$ as the reduced graph, they effectively circumvent the computational demands associated with calculating node embeddings for large graphs.
Similarly, CoreWalk~\cite{brandeis2020kcorewalk} utilizes this framework for reducing the graph but generate the node embeddings with different methods.
Furthermore, recent work WIS~\cite{razin2023WIS} highlights that the ability to model node interactions is primarily determined by the model-free metrics \textit{walk index} (WI), which refers to the number of walks originating from the boundary of the partition. Consequently, unimportant edges are removed based on sorted WI values. 
Although these metrics offer insights from a certain perspective of the graph, they might not be compatible with the downstream models and tasks.

In contrast, recent years have also witnessed many model-based scoring methods, which utilize a parameterized model to calculate the score. For instance, \citet{jin2021gcond} adopt coreset methods~\citep{welling2009herding,sener2018activecoreset} to select nodes based on their embeddings from a trained GNN model. 
Apart from these general scoring methods which can be used for any modality, recent works on the interpretability of GNN, e.g., \textit{GNNexplainer}~\cite{ying2019gnnexplainer} can also be related to graph sparsification. 
For example, IGS~\cite{braingnn-igs} sparsifies a graph based on edge importance obtained from \textit{GNNexplainer} and feeds the sparsified graph into the next iteration. 
Other methods fall under graph structure learning~\cite{zhu2021surveygsl}, which can be viewed as renders the scoring of edges learnable.  For example, \citet{zheng2020neuralsparse} learn a sparsified graph structure by neighbor sampling according to the and reparameterization trick~\cite{jang2016repara} during GNN training; UGS~\cite{chen2021sparseloottery} and CGP~\citep{liu2023cgp} simultaneously prune the elements in graph adjacency matrix (edges) and the GNN weights to reduce the graph and lighten the model. Recognizing that UGS fails to preserve topology, GST \cite{zhang2024two} is proposed to combine semantic and topological information during sparsification. 
Differing from the iterative approaches described above, TEDDY~\cite{seo2024teddy} introduces a one-shot reduction strategy. This strategy begins by sparsifying the graph using only degree information, followed by training a lightweight model through knowledge distillation and Projected Gradient Descent (PGD) on the $l_0$ ball. However, these structure learning methods are closely tied to specific selection models or sparsified GNN models. In other words, these graphs are merely byproducts of graph structure learning, even though they have the potential to serve as reduced inputs for other algorithms. The transferability and universality of their sparsified graphs to other graph algorithms remain inadequately explored.

There are a few works that do not fall into the scoring-and-selection strategy and instead adopt a holistic approach to selection. For example, OTC~\cite{garg2019solving} achieves selection by minimizing the optimal transport distance between an initial node distribution supported on $\mathcal{V}$ and a target distribution supported on a subset $\mathcal{V}'$ selected via the associated transport problem~\cite{peyre2019computational}. The reduced graph is then obtained by restricting the original graph to nodes in $\mathcal{V}'$ and their incident edges.

\subsection{Graph Coarsening}\label{subsec:graphcoarsening}
The selection of nodes or edges in sparsification methods can inevitably lose some information. To ensure that a sufficient amount of information is preserved, coarsening techniques have been developed, which involve grouping nodes and aggregating them.
This process can be carried out iteratively, yielding hierarchical views of the original graph. Existing coarsening methods can be categorized into two groups depending on whether a reconstruction objective exists: reconstruction-based methods and reconstruction-free methods, which will be elaborated upon subsequently.



\subsubsection{Reconstruction-Based Methods}

Reconstruction-based coarsening methods involve a two-step process. First, they reconstruct the original graph from the coarse graph, where super nodes are mapped back to their original nodes. This way, the super nodes are \textit{lifted} to sizes comparable to those in the original graph~\citep{lefevre2010grass1}.  Subsequently, the goal is to find the coarsening mapping matrix (${\bf C}$ or ${\bf P}$) that can minimize the differences between the reconstructed graph and the original one, which are quantified by examining their adjacency or Laplacian matrices. 
These coarsening techniques can be broadly categorized into spatial or spectral coarsening methods, depending on whether they utilize the adjacency or Laplacian matrix for this purpose.

\textit{Spatial coarsening.} Spatial coarsening adopts the Reconstruction Error (RE) \citep{lefevre2010grass1} as the objective function $\mathcal{L}$:
\begin{equation}
    RE_p({\bf A}_l|{\bf A})=||{\bf A}_l-{\bf A}||^p_\text{F}
\end{equation}
where the \textit{lifted} adjacency matrix ${\bf A}_l$~\citep{lefevre2010grass1} is usually defined as:
\begin{equation}
{\bf A}_l(u,v)=\begin{cases}0&\mathrm{~if~}u=v\\
{E_i}/{\binom{N_i}2}&\mathrm{~if~}u,v\in V_i\\
{E_{ij}}/{(N_iN_j)}&\mathrm{~if~}u\in V_i,v\in V_j\end{cases}
\end{equation}
where $E_i$ represents the number of edges within the super node $V_i$, $E_{ij}$ denotes the number of edges between $V_i$ and $V_j$, and $N_i$ is the number of nodes belonging to $V_i$.
It is also proved that ${\bf A}_l$ can be expressed as a function of $\bf P$ and $\bf A$~\cite{riondato2017S2l}.
As the first work proposing RE, GraSS~\citep{lefevre2010grass1} randomly samples part of node pairs and merges one of them causing the smallest increase of RE. 
~\citet{riondato2017S2l} show the connection of minimizing RE with 
geometric clustering problems and develops a polynomial-time approximation. Similarly,~\citet{beg2018scalableapproximation} propose a weighted sampling scheme to sample vertices for merging that will result in the least RE. 

\textit{Spectral coarsening.} 
Different from spatial methods, spectral coarsening methods compare the ${\bf L}_l$ and $\bf L$ by comparing their eigenvalues or eigenvectors. The \textit{lifted} Laplacian matrix is defined as ${\bf L}_l={\bf P}^\top{\bf L}'{\bf P}$~\citep{kumar2023fgc1}.
\citet{loukas2019spectral&cut,loukas2018spectrally} propose \textit{restricted spectral approximation} and derive a relaxed evaluation called Relative Eigenvalue Error (REE) defined as $\operatorname{REE} = \sum_{i=1}^k|\lambda_i-\lambda'_i|/\lambda_i$
, where $\lambda_i$ and $\lambda'_i$ are the top-$k$ eigenvalues of the matrices ${\bf L}$ and ${\bf L}'$, respectively. Note that they use ${\bf L}'$ instead of ${\bf L}_l$ because the comparison of eigenvalues does not require the alignment of the sizes.
They also give the theoretical guarantee of \textit{greedy pairwise contraction} approaches, where different node pair scoring methods can be used including Heavy Edge~\citep{dhillon2007heavy}, Algebraic Distance~\citep{chen2011algebraicad}, Affinity~\citep{livne2012affinity} and Local Variation (LV)~\citep{loukas2019spectral&cut}. 
Some works hold that the edge weights can be further optimized after pairwise contraction. For example, 
~\citet{zhao2018nearlylinearvisualization} scale the edge weights by stochastic gradient descent to further align the eigenvalues after coarsening.  In addition, some endeavor is made for lossless coarsening, e.g.,~\citet{navlakha2008lossless1,khan2015lossless2} keep the correction set recording the missed edges during pairwise contraction. 
 
Aside from these heuristics, there are other approaches. FGC~\citep{kumar2023fgc1} takes both the graph structure and the node attributes as the input and alternatively optimizes $\bf C$ and ${\bf X}'$. 
SGC~\citep{bravo2019unifyingsparse&coarse} considers the edge sparsification and contraction as edge weights of 0 and $\infty$. Then they develop a
probabilistic framework to preserve the pseudo-inverse of graph Laplacian ${\bf L}^+$ by ${\bf L}_l^+$. GOREN~\citep{cai2021goren} learns the edge weights in the coarse graph by a GNN with the loss to preserve the graph Laplacian $\bf L$ by ${\bf L}_l$.

\subsubsection{Reconstruction-Free Methods}
Despite the proliferation of reconstruction-based methods, other approaches do not rely on the reconstruction while still keeping the key information. 
~\cite{itzkovitz2005coarsecomplexnetworks} coarsen complex engineered and biological networks into smaller and more comprehensible versions, where nodes represents probabilistically generalized network motifs in the original network. 
To analyze social networks with diverse attributes and relations, SNAP~\citep{tian2008ksnap} produces a summary graph where every node inside a super node has the same values for selected attributes, and is adjacent to nodes with the same selected relations. \textit{k}-snap~\citep{tian2008ksnap} relaxes this homogeneity requirement and allows users to control the resolutions of summaries.
AGSUMMARY~\citep{wu2014ag} utilizes the \textit{Minimum Description Length} principle to design a cost function and compute an optimal summary by
neighborhood greedy strategy. Since the former two methods only apply to discrete attributes, CANCEL~\cite{zhang2010discoverycancel} relaxes this condition to continual ones with adaptive cutoffs and proposes a comprehensive metric named~\textit{interestingness}.
To extend the above methods focusing only on one task, CoarseNet~\cite{purohit2014coarsenet} tries to find a succinct representation of the original network that preserves important diffusive characters, which can be applied to both influence maximization and  propagation patterns detection tasks. To flexibly achieve higher performance among different tasks, Netgist~\cite{amiri2018netgist} defines a task-based graph summarization problem and uses RL to create a flexible framework for learnable node merging policies.

\subsubsection{Remarks on Graph Coarsening}
\textit{Graph Coarsening in GNNs.} There are growing numbers of works that combine coarsening with GNNs.
To mimic the pooling layer in the convolutional neural network, \citet{such2017coarsepool} make the mapping matrix learnable and produce a pooled graph reduced to fewer nodes layer by layer. 
$k$-MIS~\cite{bacciu2023downsamplng} introduces a
unifying interpretation of pooling in both regular and graph data with a controllable equispaced graph coarsening mechanism. 
Borrowing from graph covers and k-plexes (pseudo-cliques where each node can miss up to 
$k$ links), KPlexPool~\citet{bacciu2021kplex} is developed as a non-parametric pooling technique that can generalize effectively to graphs with varying natures and connectivity patterns.

While coarsening methods for pooling are mostly tailored for graph classification, recent years have also seen the development of coarsening methods for node classification.
For instance, SCAL~\citep{huang2021scaling} first trains a GNN model in a graph coarsened by LV, with super node label defined as ${\bf Y}'=\operatorname{argmax}({\bf P}{\bf Y})$ and then directly uses this model to inference. 
Following the same framework, \citet{generale2022scalingrgcn} define the relation summarization and uses R-GCN~\cite{schlichtkrull2018rgcn} for knowledge graphs.
CONVMATCH~\cite{dickens2023convolutionmatch} merges nodes that are equivalent or similar w.r.t. the GCN convolution operation. Similarly, ~\citet{buffelli2022sizeshiftreg} match the node embeddings output by GNNs among graphs in different coarsening ratios to deal with one of data shift problem --- size shift.

Some works employ coarsening techniques to improve the scalability when training GNNs in large-scale graphs.
VNG~\citep{si2022serving} highlights the ongoing challenge of efficiently deploying a GNN model in online applications when edges exist between testing nodes and training nodes. To tackle this issue, they match the forward propagation to obtain the mapping matrix ${\bf C}$ by applying weighted k-means.
Recognizing that only a subset of nodes (target nodes) requires analysis in large-scale web graphs, \textit{Graph Skeleton}~\cite{cao2024graph} focuses reducing the background nodes while preserving the classification performance of target nodes. It merges background nodes sharing the same distance to target nodes, as well as target nodes and their affiliated nodes---that is, background nodes linked exclusively to a single target node. 

Some methods do not change the node features but the paths between each pair of target nodes. For example, Kron reduction~\citep{dorfler2012kron} is initially developed to address challenges in electrical networks, specifically to simplify resistance networks while maintaining the pairwise ER.
This method calculates the coarse graph Laplacian ${\bf L'}$ by \textit{Schur complement}
\begin{equation}
{\bf L'}={\bf L}_{\mathcal{V}',\mathcal{V}'}-{\bf L}_{\mathcal{V}',\bar {\mathcal{V}'}}{\bf L}_{\bar {\mathcal{V}'},\bar {\mathcal{V}'}}^{-1}{\bf L}_{\bar {\mathcal{V}'},\mathcal{V}}
\end{equation}
where $\mathcal{V}'$ denotes the selected index from $\mathcal{V}$, $\bar {\mathcal{V}'}=\mathcal{V}-\mathcal{V}'$, and ${\bf L}_{\mathcal{A},\mathcal{B}}$ is the submatrix of ${\bf L}$ whose row index is $\mathcal{A}$ and column index is $\mathcal{B}$. Recently, ~\citet{sugiyama2023krondirected} extend it to a directed graph with self-loop. In contrast to selecting nodes arbitrarily, ~\citet{fang2010multilevelindependentedge} calculate \textit{Schur complement} after finding the largest node set consisting of nodes not adjacent to each other.



\textit{Connections with Graph Clustering/Partition.}
Partition and clustering in graphs are long-developed areas. Graph partition aims to find a split for a graph with the least cost, e.g., cutting the fewest edges. The representative partition method, METIS~\cite{karypis1997metis}, coarsens a graph iteratively by pairwise contraction, splits the nodes in the coarsest graph, and reversely maps them to the original graph. This reconstruction-based framework is widely used for partition~\cite{safro2015advancedcoarsenpartition}, which means the areas of partition and coarsening are mutually reinforcing.
Graph clustering attempts to find groups of nodes with high in-group edge density, and relatively
low out-group density~\cite{tsitsulin2023graphclustegnn}. Thus, by mapping these dense groups to super nodes and aggregating them, any clustering method can apply to the coarsening strategy.





\begin{table*} [t] 
\caption{Summary of representative graph reduction methods. NC -- Node Classification, GC -- Graph Classification, LP -- Link Prediction, and AD -- Anomaly Detection. The ``Input'' column shows the type of the input graph. $\tilde{{\bf A}}$ indicates that the method can only applied to a symmetric adjacency matrix. ${\bf A}_r$ denotes the adjacency matrix with multiple relations.}
\label{tab:taxonomy}
\renewcommand{\arraystretch}{1.15}
\centering
\small
\begin{tabular}{lllll} 
\toprule
\textbf{\textbf{Approach}} & \textbf{\textbf{Method}} & \textbf{\textbf{Learning Objective}} & \textbf{\textbf{Evaluation}} & \textbf{\textbf{Input}}  \\  \hline
\multirow{6}{*}{Graph Sparsification} & SPANNER~\cite{althofer1993spannerprove} & \multirow{3}{*}{Property Preservation} & Value of $\epsilon$ & $\tilde{{\bf A}}$ \\ \cline{2-2} \cline{4-5} 
 & TRS~\cite{batson2009twice} & & Value of $\epsilon$ & $\tilde{{\bf A}}$ \\  \cline{2-2} \cline{4-5} 
 & SparRL~\cite{wickman2022sparsebyRL} & & Value of $\epsilon$ & $\tilde{{\bf A}}$ \\  \cline{2-5}
 & WIS~\cite{razin2023WIS} & \multirow{3}{*}{Performance Preservation} & NC &  ${\bf X}, {\bf A}, {\bf Y}$\\  \cline{2-2} \cline{4-5} 
 & CoreWalk~\cite{brandeis2020kcorewalk} & & LP & ${\bf A}$ \\  \cline{2-2} \cline{4-5} 
 & UGS~\cite{chen2021sparseloottery} & & NC, LP & ${\bf X}, {\bf A}, {\bf Y}$ \\ \hline 
\multirow{6}{*}{Graph Coarsening} & \textit{k}-snap~\cite{tian2008ksnap} & \multirow{4}{*}{Reconstruction-Free} & Clustering Metrics & ${\bf X}, {\bf A}_r$  \\ \cline{2-2} \cline{4-5}
 & NetGist~\cite{amiri2018netgist} & & Clustering Metrics & ${\bf A}$ \\ \cline{2-2} \cline{4-5}
 & SCAL~\cite{huang2021scaling} & & NC & ${\bf X}, {\bf A}$ \\ \cline{2-2} \cline{4-5}
  & VNG~\cite{si2022serving} & & NC & ${\bf X}, {\bf A}$\\ \cline{2-5}
 & LV~\cite{loukas2018spectrally} & \multirow{2}{*}{Reconstruction-Based} & REE & ${\bf A}$ \\ \cline{2-2} \cline{4-5} 
 & FGC~\cite{kumar2023fgc1} & & REE, RE, NC &  ${\bf X}$, ${\bf A}$ \\  \hline

\multirow{21}{*}{Graph Condensation} & GCond \citep{jin2021gcond} & \multirow{11}{*}{Gradient Matching} & NC & \textbf{X}, \textbf{A}, \textbf{Y} \\ \cline{2-2} \cline{4-5} 
 & DosCond \citep{jin2022doscondensing} &  & NC, GC & \textbf{X}, \textbf{A}, \textbf{Y} \\ \cline{2-2} \cline{4-5} 
 & MSGC \citep{gao2023multiple} &  & NC & \textbf{X}, \textbf{A}, \textbf{Y} \\ \cline{2-2} \cline{4-5} 
 & SGDD \citep{yang2023does} &  & NC, LP, AD & \textbf{X}, \textbf{A}, \textbf{Y} \\ \cline{2-2} \cline{4-5} 
 & FGD \citep{feng2023fair} & & NC & \textbf{X}, \textbf{A}, \textbf{Y} \\ \cline{2-2} \cline{4-5} 
 & GCARe \citep{mao2023gcare} &  & NC & \textbf{X}, \textbf{A}, \textbf{Y} \\ \cline{2-2} \cline{4-5} 
 & CTRL \citep{zhang2024two} &  & NC, GC & \textbf{X}, \textbf{A}, \textbf{Y} \\ \cline{2-2} \cline{4-5} 
 & GroC \citep{li2023attend} &  & NC & \textbf{X}, \textbf{A}, \textbf{Y} \\ \cline{2-2} \cline{4-5}
 & EXGC \citep{fangjf_exgc} &  & NC, GC & \textbf{X}, \textbf{A}, \textbf{Y} \\ \cline{2-2} \cline{4-5} 
 & FedGKD \citep{pan2023fedgkd} &  & NC & \textbf{X}, \textbf{A}, \textbf{Y} \\ \cline{2-2} \cline{4-5} 
 & MCond \citep{gao2023graph} &  & NC & \textbf{X}, \textbf{A}, \textbf{Y} \\ \cline{2-5}
 & GCDM \citep{liu2022graph} & \multirow{4}{*}{Distribution Matching} & NC & \textbf{X}, \textbf{A}, \textbf{Y} \\ \cline{2-2} \cline{4-5} 
 & PUMA \citep{liu2023puma} &  & NC & \textbf{X}, \textbf{A}, \textbf{Y} \\ \cline{2-2} \cline{4-5} 
 & CaT \citep{liu2023cat} &  & NC & \textbf{X}, \textbf{A}, \textbf{Y} \\ \cline{2-2} \cline{4-5} 
 & DisCo \citep{xiao2024disentangled} &  & NC & \textbf{X}, \textbf{A}, \textbf{Y} \\ \cline{2-5} 
 & KiDD \citep{xu2023kernel} & \multirow{2}{*}{Kernel Ridge Regression} & GC & \textbf{X}, \textbf{A}, \textbf{Y} \\ \cline{2-2} \cline{4-5} 
 & GC-SNTK \citep{wang2023fast} &  & NC & \textbf{X}, \textbf{A}, \textbf{Y} \\ \cline{2-5} 
  & GCEM \citep{liu2023graph} & Eigenbasis Matching & NC & \textbf{X}, \textbf{A}, \textbf{Y} \\ \cline{2-5} 
 & MIRAGE \citep{gupta2023mirage} & Computation Tree Compression & GC & \textbf{X}, \textbf{A}, \textbf{Y} \\ \cline{2-5}
  & SFGC \citep{zheng2023structure} & \multirow{2}{*}{Trajectory Matching} & NC & \textbf{X}, \textbf{A}, \textbf{Y} \\ \cline{2-2} \cline{4-5} 
 & GEOM \citep{zhang2024Navigating} &  & NC & \textbf{X}, \textbf{A}, \textbf{Y} \\ 
 \bottomrule
\end{tabular}
\end{table*}

\subsection{Graph Condensation} \label{graphCondensation}
While sparsification and coarsening methods have proven effective in reducing the size of graph data, they have inherent limitations. 
As many of these methods prioritize preserving specific graph properties, they do not leverage the downstream task information and could lead to suboptimal model performance.
Furthermore, these techniques rely on the assumption of the existence of representative nodes or edges in the original graph, which might not always hold true in the original dataset. To address these issues, graph condensation, first introduced by~\cite{jin2021gcond}, has come into play.

Motivated by dataset  distillation~\cite{wang2018dataset} and dataset condensation~\cite{zhao2020dataset}, graph condensation revolves around condensing knowledge from a large-scale graph dataset to construct a much smaller synthetic graph from scratch. The goal is to ensure that models trained on this condensed graph dataset exhibit comparable performance to those trained on the original one. In other words, we can see graph condensation as a process of minimizing the loss defined on the models trained on the real graph $\mathcal{T}$ and the synthetic graph $\mathcal{S}$. Thus, the objective function in Eq.~\eqref{eq:graphReduction} can reformulated as follows: 
\begin{equation}
\mathcal{S}=\underset{\mathcal{S}}{\arg \min } \ \mathcal{L}(\operatorname{GNN}_{\boldsymbol{\theta}_{\mathcal{S}}}(\mathcal{T}), \operatorname{GNN}_{\boldsymbol{\theta}_{\mathcal{T}}}(\mathcal{T})), \label{eq:gcondL}
\end{equation}
where  $\operatorname{GNN}_{\boldsymbol{\theta}_{\mathcal{S}}}$ and $\operatorname{GNN}_{\boldsymbol{\theta}_{\mathcal{T}}}$ denote the GNN models trained on $\mathcal{S}$ and $\mathcal{T}$, respectively; $\mathcal{L}$ represents the loss function used to measure the difference of these two models. Based on the specific designs of $\mathcal{L}$, we classify existing graph condensation methods into three categories: matching-based methods, kernel ridge regression methods, and others.

\subsubsection{Matching-Based Methods}
To find the optimum synthetic graph dataset that minimizes the loss for a GNN trained on it, while having the lowest loss on the original graph dataset, one approach is to match some meta-data elements related to $\mathcal{S}$ and $\mathcal{T}$ like gradients w.r.t. the model parameters and distribution of node classes.

\textit{Gradient Matching.} 
For computing the optimum synthetic graph dataset $\mathcal{S}$, Eq. \eqref{eq:gcondL} can be rewritten as the following bi-level problem that generalizes to the distribution of random initialization $P_{\boldsymbol{\theta}_0}$:

\begin{subequations}
    \begin{equation}
\min _{\mathcal{S}} \mathrm{E}_{\boldsymbol{\theta}_0 \sim P_{\boldsymbol{\theta}_0}}\left[\mathcal{L}\left(\operatorname{GNN}_{\boldsymbol{\theta}_{\mathcal{S}}}(\mathbf{A}, \mathbf{X}), \mathbf{Y}\right)\right], \label{eq:bilevel} \quad 
\end{equation}
\begin{equation}
\\
\text { s.t. } \quad \boldsymbol{\theta}_{\mathcal{S}}=\underset{\boldsymbol{\theta}}{\arg \min } \ \mathcal{L}\left(\operatorname{GNN}_{\boldsymbol{\theta}\left(\boldsymbol{\theta}_0\right)}\left(\mathbf{A}^{\prime}, \mathbf{X}^{\prime}\right), \mathbf{Y}^{\prime}\right),\label{eq:bilevel2}
\end{equation}
\end{subequations}
where $\boldsymbol{\theta}\left(\boldsymbol{\theta}_0\right)$ denotes that $\boldsymbol{\theta}$ is a function acting on $\boldsymbol{\theta}_0$. 
To simplify the bi-level optimization of Eq. \eqref{eq:bilevel} and \eqref{eq:bilevel2},~\citet{jin2021gcond} propose GCond framework, the first graph condensation method, that matches the gradients from both graph datasets match during each step of training:
\begin{subequations}
\begin{equation}\label{eq:gradMatching}
    \min _{\mathcal{S}} \mathrm{E}_{\boldsymbol{\theta}_0 \sim P_{\theta_0}}\left[\sum_{t=0}^{T-1} D\left(\nabla_{\boldsymbol{\theta}} \mathcal{L}_1, \nabla_{\boldsymbol{\theta}} \mathcal{L}_2\right)\right], 
\end{equation}
\begin{equation}
    \mathcal{L}_1 = \mathcal{L}\left(\mathrm{GNN}_{\boldsymbol{\theta}_t}\left(\mathbf{A}^{\prime}, \mathbf{X}^{\prime}\right), \mathbf{Y}^{\prime}\right), \label{eq:L1}
\end{equation}
\begin{equation}
    \mathcal{L}_2 = \mathcal{L}\left(\mathrm{GNN}_{\boldsymbol{\theta}_t}\left(\mathbf{A}, \mathbf{X}\right), \mathbf{Y}\right),
\end{equation}
\end{subequations}
where $D(\cdot,\cdot)$ represents a distance function, $T$ stands for the total number of steps in the entire training trajectory, and $\boldsymbol{\theta}_t$ refers to the model parameters at $t$-th training epoch. By optimizing the above objective, the training process on the smaller synthetic graph dataset $\mathcal{S}$ mimics the path taken on the larger real dataset $\mathcal{T}$, which leads to models trained on real and synthetic datasets ending up with similar solutions.
To prevent overlooking the implicit correlations between node attributes and graph structure, GCond condenses the graph structure by leveraging a function to parameterize the adjacency matrix $\mathbf{A'}$:
\begin{equation}
    \mathbf{A}_{i j}^{\prime}=\sigma \left(\left[{\operatorname{MLP}_{\Phi}\left(\left[\mathbf{x}_i^{\prime} ; \mathbf{x}_j^{\prime}\right]\right)+\operatorname{MLP}_{\Phi}\left(\left[\mathbf{x}_j^{\prime} ; \mathbf{x}_i^{\prime}\right]\right)}\right]/{2}\right)    
\end{equation}
where $\operatorname{MLP}_{\Phi}$ is a multi-layer perceptron (MLP) parameterized with ${\Phi}$ and $[.;.]$ indicates concatenation. However, the optimization process in GCond involves a nested loop as shown in Eq.~\eqref{eq:gradMatching}, which hinders the scalability of the condensation method. 
To address this, DosCond~\cite{jin2022doscondensing} proposes a one-step GM scheme, where it exclusively matches the network gradients for the model initialization $\boldsymbol{\theta}_{0}$ while discarding the training trajectory of $\boldsymbol{\theta}_{t}$. By dropping the summation in Eq.~\eqref{eq:gradMatching}, the objective function of DosCond becomes:
\begin{subequations}
\begin{equation}
    \min _{\mathcal{S}} {\mathrm{E}_{\boldsymbol{\theta}_0 \sim P_{\boldsymbol{\theta}_0}}}\left[D\left(\nabla_\theta \mathcal{L}_1, \nabla_\theta \mathcal{L}_2\right)\right].
\end{equation}
\end{subequations}
Note that, DosCond treats the graph structure ${\bf A}'$ as a probabilistic model to learn a discretized graph structure by learning a Bernoulli distribution over the edge set.
Moreover, DosCond offers \textbf{a theoretical insight} into the GM scheme in graph condensation:  the smallest gap between the resulting loss (achieved by training on synthetic graphs) and the optimal loss is upper bounded by the gradient matching loss. Additionally, it is worth mentioning that DosCond is the first method that does graph condensation focusing on graph classification for reducing the number of multiple graphs.  In subsequent research,  EXGC~\cite{fangjf_exgc} further identifies two primary causes for the inefficiency of those graph condensation methods: the concurrent updating of large parameter sets and the parameter redundancy. Built on the GM scheme, it employs the Mean-Field variational approximation to expedite convergence and integrate explanation techniques~\cite{ying2019gnnexplainer} to selectively focus on important nodes during the training process, thereby enhancing the efficiency of graph condensation.

Several subsequent studies target at improving GM for graph condensation to enhance the effectiveness of GCond~\citep{gao2023multiple,yang2023does,feng2023fair,mao2023gcare,li2023attend,gao2023graph,zhang2024two}. Unlike GCond, which uses a single fully connected graph to generate the condensed graph dataset $\mathcal{S}$, MSGC~\citep{gao2023multiple} is introduced to leverage multiple sparse graphs to create diverse neighborhoods for nodes that enhance the capturing of neighborhood information. This, in turn, allows GNNs to generate more informative embeddings in the condensed graphs.
Regarding the generalizability of GCond across different GNN architectures, SGDD~\citep{yang2023does} is proposed to explicitly prevent overlooking the original graph dataset structure $\mathbf{A}$ by broadcasting it into the construction of synthetic graph structure $\mathbf{A'}$. In this way, it is shown that SGDD reduces the Laplacian Energy Distribution (LED) \citep{das2016distribution,gutman2006laplacian} shift crossing various datasets significantly compared to GCond.  In addition to the node classification task, to validate the effectiveness of SGDD, extensive link prediction problems have been explored. 
\citet{gao2023graph} identify the potential issues in existing graph condensation methods for inductive node representation learning and emphasize the under-explored need for an explicit mapping between original and synthetic nodes. Consequently, a GM-based method named MCond is introduced, which explicitly learns a sparse mapping matrix to smoothly integrate new nodes into the synthetic graph for inductive representation learning. MCond employs an alternating optimization scheme compared to GCond, allowing the synthetic graph and mapping matrix to take turns updating toward dedicated objectives. Furthermore, CTRL \citep{zhang2024two} argues the limited approach of using cosine similarity for gradient matching, leading to biases, and suggests adding gradient magnitudes into the objective function introduced in GCond for a more accurate match. Their empirical findings also show that this approach better aligns frequency distributions between condensed and original graphs. 

Despite the effectiveness of the previously mentioned graph condensation methods,
\citet{feng2023fair} recognize that these methods tend to exhibit fairness issues. By identifying the group fairness\footnote{Group fairness in algorithms ensures unbiased and fair treatment across diverse demographic groups. It seeks to prevent any form of discrimination or bias against specific groups within the algorithmic decision-making process~\citep{mehrabi2021survey}}, it demonstrated that as distillation performance increases, fairness (Demographic Parity $\Delta_{D P}$) decreases \citep{feng2023fair}. Particularly, it is showcased that, by measuring the fairness of GNNs trained on original graphs versus those trained on condensed graphs, an improvement in performance correlates with heightened fairness issues in the synthetic condensed graph. To address this challenge, FGD is introduced, as a fair graph condensation method. This is achieved by incorporating the coherence metric into the GM loss function outlined in Eq.~\eqref{eq:gradMatching}. Particularly, the coherence metric is a bias calculator that captures the variance of the estimated sensitive group membership. 
Similarly, to address the fairness issue of current graph condensation methods,~\citet{mao2023gcare} propose graph condensation with Adversarial Regularization (GCARe), which is a method that directly regularizes the condensation process to distill the knowledge of different subgroups fairly into resulting graphs. Also, other studies, such as FedGKD \citep{pan2023fedgkd}, utilize gradient matching-based graph condensation in federated graph learning. 

\textit{Distribution Matching.} 
While GM-based methods offer benefits compared to traditional methods, it faces two challenges. 
\textbf{First}, the condensation process becomes computationally expensive when minimizing the GM loss due to the need for computing second-order derivatives with respect to GNN parameters. 
\textbf{Second}, the architecture-dependent nature of the GM loss may hinder the condensed graph's generalization to new GNN architectures~\citep{liu2022graph}.
Alternatively, the Distribution Matching (DM) approach seeks to acquire synthetic graph data whose distribution closely approximates that of real data.
To address the limitations of GM-based methods, such as the reduced generalizability of graph condensation across different GNN architectures and the computational overhead, DM-based algorithms directly optimize the distance between the two distributions using metrics such as Maximum Mean Discrepancy (MMD)~\citep{zhao2023dataset}. 
For example, CaT \citep{liu2023cat} updates the condensed graph $\mathcal{S}$ using the DM objective function to find the optimal synthetic graph as follows: 
\begin{equation}
    \ell_{\mathrm{MMD}}=\sum_{c \in \mathcal{C}} r_c \cdot\left\|\operatorname{Mean}\left(\boldsymbol{E}_{c}\right)-\operatorname{Mean}\left(\tilde{\boldsymbol{E}}_{c}\right)\right\|^2,
\end{equation}
where ${\mathcal{C}}$ is the set of node classes, ${\boldsymbol{E}}_{c}$ and $\tilde{\boldsymbol{E}}_{c}$ are the embeddings of nodes with class $c$ in the original and condensed graph, respectively, and $r_c$ is the class ratio for class~$c$. Likewise, GCDM~\citep{liu2022graph} regards the original graph as a distribution of receptive fields and seeks to synthesize a smaller graph whose receptive fields share a similar distribution to that of the original graph. Other works, such as PUMA~\citep{liu2023puma}, employ a similar approach for various applications like continual learning.

DisCo~\citep{xiao2024disentangled} addresses scalability issues in current matching-based condensation methods through an iterative process that condenses nodes and edges separately. In the node condensation step, synthetic nodes are generated using an MLP pre-trained on $\mathcal{T}$ while considering label distribution. For the synthetic graph structure, edges are predicted using a pre-trained link prediction model, matching the original graph's edge distribution. Notably, DisCo is found to be significantly faster than existing methods because it conducts separate condensation processes for edges and nodes.

\subsubsection{Kernel Ridge Regression Methods} 
To mitigate heavy computation in the optimization problem in Eq.~\eqref{eq:bilevel}, KIDD~\citep{xu2023kernel}, the first Kernel Ridge Regression (KRR) method for graph condensation, simplifies the optimization objective into a single-level problem by substituting the closed-form solution of the lower-level problem into the upper-level objective.
To implement KRR for graph-level tasks, a graph kernel is essential~\citep{xu2023kernel}. Thus, a Graph Neural Tangent Kernel (GNTK) ~\citep{du2019graph} for the KRR graph classifier is chosen, as GNTK effectively characterizes the training dynamics of GNNs and yields such a closed-form solution. Concretely, if $\operatorname{GNN}_{\boldsymbol{\theta}_{\mathcal{S}}}$ in Eq.~\eqref{eq:bilevel} is instantiated as the KRR and the squared loss is applied, Eq.~\eqref{eq:bilevel} and Eq.~\eqref{eq:bilevel2} can be instantiated as a single objective function which is as follows:
\begin{subequations}
    \begin{equation}
        \min _{\mathcal{S}} \mathcal{L}_{\mathrm{KRR}}=\frac{1}{2}\left\|\mathbf{y}_{\mathcal{T}}-\mathbf{K}_{\mathcal{T} \mathcal{S}}\left(\mathbf{K}_{\mathcal{S} \mathcal{S}}+\epsilon \mathrm{I}\right)^{-1} \mathbf{y}_{\mathcal{S}}\right\|^2,
    \end{equation}
\end{subequations}
where $\epsilon > 0$ is a KRR hyper-parameter, $\mathbf{K}_{\mathcal{TS}}$ is the kernel matrix between original and synthetic graphs while $\mathbf{K}_{\mathcal{SS}}$ is the kernel matrix between synthetic graphs\footnote{Each kernel indicates infinitely wide multi-layer GNNs trained by gradient descent through squared loss \citep{du2019graph}}.
Also, $\mathbf{y}_{\mathcal{S}}$ and $\mathbf{y}_{\mathcal{T}}$ are the concatenated graph labels from real dataset and synthetic dataset, respectively.

\subsubsection{Other Methods} 
Most graph condensation methods involve GNNs or graph filters when generating condensed graphs, which can be biased to a specific spectrum and potentially miss the overall distribution of the real graph~\citep{liu2023graph}. To overcome this, \citet{liu2023graph} propose to avoid spectrum bias in the condensation process, which is caused by utilizing GNNs. To obtain representation spaces similar to the ones in the real graph, GCEM \citep{liu2023graph} matches the representative eigenbasis (the underlying graph structure) of real and synthetic graphs during condensation.
Nevertheless, due to the differing sizes of the subspaces defined by the eigenvectors of the real and synthetic graphs, direct alignment is not feasible, prompting GCEM to match the node attributes in the subspaces as an alternative, which will make them share similar distributions:
\begin{equation}
    \mathcal{L}_e=\sum_{c=1}^C \sum_{k=1}^K\left\|\overline{\mathbf{h}}_{c, k}-\overline{\mathbf{h}^{\prime}}{ }_{c, k}\right\|^2,
\end{equation}
where ${\mathbf{h}}_{c, k}$ and ${\mathbf{h'}}_{c, k}$ are the representation of the $c$-th class center in $k$-th subspace for real and synthetic graphs, respectively.

\citet{gupta2023mirage} investigate that in GM-based methods, the gradients of model weights are contingent on various factors such as the specific GNN architecture and hyper-parameters. This leads to a reduction in performance when alternating the GNN during testing. Furthermore, the requirement for the original graph for training in graph condensation still exists, causing computational and storage constraints. With this motivation in mind, MIRAGE \citep{gupta2023mirage} is introduced to condense multiple graphs to address graph classification problems. It utilizes GNNs to break down any graph into a collection of computation trees and then extracts frequently co-occurring computation trees from this set. It is shown that a concise set of top-k frequently co-occurring trees can effectively capture a significant portion of the distribution mass while preserving rich information. Consequently, a GNN trained solely on the frequent tree sets should be adequate for subsequent predictive tasks.

Also, \citet{zheng2023structure} propose SFGC, a structure-free graph condensation method using a matching-based approach that only outputs the condensed node features $\mathbf{X}'$, as the structure information of the real graphs is embedded in  ${\bf X}'$. Concretely, unlike gradient matching-based methods, SFGC aligns their long-term GNN training trajectories using an offline expert parameter distribution as guidance. Beyond the above efforts,  GEOM~\citep{zhang2024Navigating} makes the first attempt toward \textit{lossless graph condensation}, i.e., significantly reducing the graph size without any loss in performance. It points out that SFGC~\cite{zheng2023structure} provides biased and limited supervisory signals to difficulty nodes while overlooking easy nodes. To address this, GEOM introduces an expanding window technique to adjust the matching range for difficult and easy nodes, during the process of matching the training trajectories between $\mathcal{T}$ and $\mathcal{S}$. Remarkably, it achieves lossless graph condensation across standard benchmark datasets.  


\section{Applications} \label{sec:app}

While the primary purpose of graph reduction was to enhance the efficiency of graph algorithms, its versatility has led to its advantageous utilization in a range of applications, as will be elaborated upon in this section.


\subsection{Neural Architecture Search}

Neural architecture search (NAS)~\cite{elsken2019neural, ren2021comprehensive} focuses on identifying the most effective architecture from a vast pool of potential models to enhance generalization in a given dataset. This technique is characterized by its intensive computational demands, necessitating the training of numerous architectures on the full dataset and choosing the top performer based on validation results.
To address the computational challenge in NAS for GNNs, graph condensation methods are utilized for searching the best GNN architecture~\citep{jin2021gcond,yang2023does}. 
Specifically, the architectures are trained on the condensed graph which leads to significant speedup in the search process, and a reliable correlation in performance between training on the condensed dataset and the whole dataset is observed. Moreover,
\citet{ding2022faster} introduce a dedicated graph condensation method for NAS, highlighting that traditional graph condensation objectives fall short of achieving this objective due to their lack of generalization across GNNs. Particularly, it proposes a condensation objective for preserving the outcome of hyperparameter optimization and outperforms other graph condensation methods in terms of finding the optimal architecture.

\subsection{Continual Graph Learning}
Continual learning~\ \citep{de2021continual} aims to learn on a stream of data from a stationary data distribution, which requires tackling the issue of catastrophic forgetting, i.e., new data can interfere with current knowledge and erase it. The common strategy is \textit{memory replay}~\cite{kemker2018replay}, which involves storing representative samples from previous tasks in a buffer to recall knowledge when needed. In the context of graphs, continual learning can be benefited by informative reduced graphs. 
As introduced in Section~\ref{graphCondensation}
CaT~\citep{liu2023cat} is applied to continual graph learning by condensing the incoming graph and updating the model with condensed graphs, not the whole incoming graph. To further improve CaT,  ~\citet{liu2023puma} introduce PUMA 
which utilizes pseudo-labeling to incorporate data from unlabeled nodes, boosting the informativeness of the memory bank and addressing the problem of neglected unlabeled nodes.
In addition, the sparsification method~\cite{zhang2023ricci} 
reduces the number of nodes and edges according to the Ricci curvature evaluation and stores them into a replay buffer for continual learning.

\subsection{Visualization \& Explanation}
The reduced dataset is not only more accessible for algorithms and computers to parse but also more friendly for people to understand, which leads to the application of visualization and explanation. 
For instance, \citet{zhao2018nearlylinearvisualization} combine spectral graph coarsening method~\citep{loukas2018spectrally} and sparsification methods~\cite{feng2016spectral} to develop a nearly linear time algorithm for multilevel graph visualization.
\textit{Pyramid Transform}~\citep{shuman2015pyramid} selects nodes corresponding to top-$k$ eigenvector repeatedly and creates a multi-resolution view of large graphs.
In addition, k-core decomposition has been widely used for graph visualization and ﬁnding speciﬁc structural characteristics of networks~\cite{alvarez2005largekcore}. 
As mentioned in Section~\ref{subsec:graphcoarsening}, some coarsening methods like \textit{k}-snap~\cite{tian2008ksnap} and CANCEL~\cite{zhang2010discoverycancel} are tailed for customized discovery, which means the user can set the granularity of the graph structure and then find perspectives they are interested in.
Graph condensation is also used for visualization and explanation. For instance, GCond~\cite{jin2021gcond} observes patterns from original data by visualizing the reduced graph;
GDM~\cite{nian2023GDMexplain} employs condensation to explain GNN behavior during the training process.

\subsection{Privacy}

It has been empirically investigated that reduced datasets offer good privacy preservation~\cite{dong2022privacyforfree}.
For example, \citet{bonchi2014identity} show random sparsification can achieve meaningful levels of anonymity while preserving features of the original graph. ~\citet{DPsparsify} consider edge privacy and provide a mechanism for sparsification of a graph that ensures differential privacy and approximates the spectrum of the original graph.
However, the trade-off between the degree of reduction and the utility, i.e., preservation of key information, always exists.
Dataset condensation can be a promising technique to solve the dilemma~\cite{dong2022privacyforfree}. 
In a federated learning framework, where client devices collaboratively contribute to model development by aggregating their updates on a central server, there exists a risk that a malicious server could infer sensitive local information from the model updates sent by these clients.
FedGKD~\cite{pan2023fedgkd} trains models on condensed local graphs within each client to mitigate the potential leakage of membership of training set~\cite{shokri2017membership}. 

\subsection{Data Augmentation}
Graph data augmentation~\cite{zhao2022graph,zhao2021data,zhou2022dataaug} is commonly used to enrich the data and improve the model performance. 
Methods for graph reduction can be employed to generate various perspectives of a graph by repeatedly applying reductions at different ratios, thereby augmenting the data for subsequent models.
For example, HARP~\citep{chen2018harp} coarsens a graph in a series of levels and then embeds the hierarchy of graphs from the coarsest one to the original.
In each step, HARP initializes node embeddings in the finer-level graph using the mapped embeddings from super nodes in the coarser level. By employing various graphs in each iteration, this method augments the training data.
Meanwhile, MILE~\cite{liang2021mile} enhances this process by substituting the \textit{random walk} employed in HARP with GNNs to improve the embedding quality and efficiency.
DistMILE~\cite{he2021distmile} advances the MILE framework by adopting high-performance computing techniques.
As a condensation method, MSGC~\cite{gao2023multiple} initializes multiple small graphs by various connection schemes and employs gradient matching to optimize them. This process results in a variety of node embeddings sets, increasing diversity and thereby augmenting the data. 

\section{Future Work} \label{sec:future}
The field of graph reduction shows considerable potential, with various algorithms already being implemented across different domains. Despite achieving notable performance, current methods in graph reduction still face several challenges and limitations. In this section, we will outline these key challenges. 

\subsection{Comprehensive Evaluation}
Despite the proliferation of graph reduction methods, a significant gap exists in the field concerning the establishment of a comprehensive evaluation methodology for these emerging approaches.
As depicted in Table~\ref{tab:taxonomy},  the prevailing focus in existing graph reduction methods has primarily revolved around their ability to preserve specific graph properties or sustain the performance of GNNs on particular downstream tasks.
On one hand, the development of novel reduction algorithms should embrace a more inclusive approach, extending to the preservation of a diverse range of graph properties like homophily~\cite{zhu2020beyond,gong2023neighborhood} and accommodating various downstream tasks, such as node classification, node clustering, link prediction, graph classification, and more.
On the other hand, there is an urgent need to broaden the scope of evaluation criteria. This expansion should encompass not only the preservation of multiple graph properties but also cater to various downstream tasks concurrently. By doing so, we can gain valuable insights into the practical utility of reduced graph datasets across different applications and domains.

\subsection{Scalability}



The majority of sparsification methods have attained linear-time complexity. Similarly, most coarsening methods adopt a local approach by emphasizing smaller neighborhood-level computations to mitigate computational overhead. Consequently, scalability is presently not a prominent concern for these sparsification and coarsening strategies.
Conversely, graph condensation methods often come with higher computational costs, involving substantial memory usage and execution time due to their intricate optimization processes. Despite recent research efforts to accelerate graph condensation, the scalability issue persists. This increased computational overhead presents two primary challenges: (1) It becomes increasingly difficult to generate an informative condensed graph with a larger size without significantly increasing computational demands. (2) Applying condensation to large-scale graphs poses computational and resource challenges that require careful consideration and resolution.
\subsection{Interpretability of Condensation Process} 

While graph condensation can itself serve as an explanation or visualization of the original graph, the challenge lies in the interpretability of the condensation process. First, since most condensation methods transform the original one-hot bag-of-words node attributes ${\bf X}$ into continuous synthetic node attributes ${\bf X}'$, it remains unclear how to interpret the acquired condensed features. Second, there is the question of how to interpret the condensed graph structure. In much of the existing research on condensation, a clear correspondence between synthetic and real nodes is often lacking, giving rise to doubts about how effectively synthetic nodes encapsulate information from their real graph counterparts. One potential approach to addressing this issue is to explore the development of a general framework for enhancing interpretability during the graph condensation process and incorporating GNN interpretability techniques into this endeavor~\citep{ying2019gnnexplainer, yuan2020xgnn}.
Furthermore, it is essential to conduct further theoretical analysis to complement and expand upon the insights presented by~\citet{jin2022doscondensing}.

\subsection{Distribution Shift}
It is a common observation that GNNs often exhibit poor generalization on a test set when there exists a disparity between the distributions of the training and test sets~\cite{wu2022handling,li2022out}. 
When GNNs are trained on reduced graphs and evaluated on graphs from the original distribution, a distribution shift may occur due to the reduction process, which eliminates substantial amounts of graph elements. However, consensus is lacking regarding the definition of graph data distribution or the selection of specific properties to represent the distribution accurately. While several condensation methods employ a specific type of distribution matching as we mentioned in Section~\ref{subsec:graphCondensation}, other measures of distribution may change after the reduction, e.g., size shift~\cite{buffelli2022sizeshiftreg}. Future graph reduction should consider the potential distribution shift issues and preserve distribution-related properties to enhance the generalization of models trained on reduced graphs.


\subsection{Robustness}
Node attributes after graph reduction risk losing fidelity, posing challenges in distinguishing them from the original graph structure. This makes them susceptible to data poisoning attacks, with injected triggers during the reduction process serving as potential backdoor vulnerabilities as it happens in other data modalities like image \citep{wang2018dataset}. To improve the distillation process against such attacks, \citet{tsilivis2022can} have combined the KIP (Kernel Including Point) \citep{nguyen2020dataset,nguyen2021dataset} method using adversarial training to improve the robustness of the distilled dataset. 
However, there is a significant gap in systematic evaluation concerning their robustness for graph modality. This oversight extends to a lack of development in both attack strategies and potential defenses tailored to reduced graph structures. It is imperative that future studies investigate these aspects, focusing on the development of methodologies to assess and enhance the robustness of reduced graphs. Exploring these directions will not only provide a deeper understanding of the vulnerabilities inherent in reduced graphs but also lead to the creation of more resilient graph reduction techniques.

\subsection{Diverse Types of Graphs}
Existing graph reduction techniques have primarily concentrated on simple non-attributed or attributed graphs. Nonetheless, graph data in real-world scenarios is increasingly intricate. This complexity is evidenced by the presence of heterophilous graphs~\cite{zheng2022graphheter}, heterogeneous graphs ~\cite{bing2023heterogeneous}, directed graphs~\cite{transitivedirected}, knowledge graphs~\cite{schlichtkrull2018rgcn}, hypergraphs~\cite{antelmi2023hypersurvey} and dynamic graphs~\cite{skarding2021dynamicsurvey}. Each graph type introduces distinct characteristics, demanding a thorough comprehension of their structures and the adoption of varied graph reduction methods. While some works have developed reduction methods for heterogeneous graphs~\cite{generale2022scalingrgcn}, directed graphs~\cite{sugiyama2023krondirected} and 3D mesh graphs~\cite{baankestad2024ising}, the other types of graphs are yet to be fully explored.
\section{Conclusion} \label{sec:conclusion}

In this paper, we offer a structured and forward-looking survey of graph reduction. We begin by establishing a formal definition of graph reduction and then develop a detailed hierarchical taxonomy that systematically organizes the diverse methodologies in this area. 
Our survey divides graph reduction techniques into three primary categories: sparsification, coarsening, and condensation.  Each of these groups represents a unique approach to reducing graph complexity while preserving essential properties.  Within each category, we systematically delve into the technical intricacies of prominent methods and highlight their practical applications in various real-world scenarios. Moreover, we shed light on the existing challenges within this domain and pinpoint potential directions for future research endeavors. Our aim is to inspire and guide upcoming studies, contributing to the ongoing evolution and advancement of graph reduction methodologies.

\bibliographystyle{named}
\bibliography{ijcai24}

\begin{thebibliography}{}

\bibitem[\protect\citeauthoryear{Aho \bgroup \em et al.\egroup }{1972}]{transitivedirected}
A.~V. Aho, M.~R. Garey, and J.~D. Ullman.
\newblock The transitive reduction of a directed graph.
\newblock {\em SIAM Journal on Computing}, 1(2):131--137, 1972.

\bibitem[\protect\citeauthoryear{Alth{\"o}fer \bgroup \em et al.\egroup }{1993}]{althofer1993spannerprove}
Ingo Alth{\"o}fer, Gautam Das, David Dobkin, Deborah Joseph, and Jos{\'e} Soares.
\newblock On sparse spanners of weighted graphs.
\newblock {\em Discrete \& Computational Geometry}, 9(1):81--100, 1993.

\bibitem[\protect\citeauthoryear{Alvarez-Hamelin \bgroup \em et al.\egroup }{2005}]{alvarez2005largekcore}
J~Alvarez-Hamelin, Luca Dall'Asta, Alain Barrat, and Alessandro Vespignani.
\newblock Large scale networks fingerprinting and visualization using the k-core decomposition.
\newblock {\em Advances in neural information processing systems}, 18, 2005.

\bibitem[\protect\citeauthoryear{Amiri \bgroup \em et al.\egroup }{2018}]{amiri2018netgist}
Sorour~E Amiri, Bijaya Adhikari, Aditya Bharadwaj, and B~Aditya Prakash.
\newblock Netgist: Learning to generate task-based network summaries.
\newblock In {\em 2018 IEEE International Conference on Data Mining (ICDM)}, pages 857--862. IEEE, 2018.

\bibitem[\protect\citeauthoryear{Anderson \bgroup \em et al.\egroup }{2014}]{anderson2014unweighted}
David~G Anderson, Ming Gu, and Christopher Melgaard.
\newblock An efficient algorithm for unweighted spectral graph sparsification.
\newblock {\em arXiv preprint arXiv:1410.4273}, 2014.

\bibitem[\protect\citeauthoryear{Antelmi \bgroup \em et al.\egroup }{2023}]{antelmi2023hypersurvey}
Alessia Antelmi, Gennaro Cordasco, Mirko Polato, Vittorio Scarano, Carmine Spagnuolo, and Dingqi Yang.
\newblock A survey on hypergraph representation learning.
\newblock {\em ACM Computing Surveys}, 56(1):1--38, 2023.

\bibitem[\protect\citeauthoryear{Arora and Upadhyay}{2019}]{DPsparsify}
Raman Arora and Jalaj Upadhyay.
\newblock On differentially private graph sparsification and applications.
\newblock In {\em Advances in Neural Information Processing Systems}, volume~32, 2019.

\bibitem[\protect\citeauthoryear{Bacciu \bgroup \em et al.\egroup }{2021}]{bacciu2021kplex}
Davide Bacciu, Alessio Conte, Roberto Grossi, Francesco Landolfi, and Andrea Marino.
\newblock K-plex cover pooling for graph neural networks.
\newblock {\em Data Mining and Knowledge Discovery}, 35(5):2200--2220, 2021.

\bibitem[\protect\citeauthoryear{Bacciu \bgroup \em et al.\egroup }{2023}]{bacciu2023downsamplng}
Davide Bacciu, Alessio Conte, and Francesco Landolfi.
\newblock Generalizing downsampling from regular data to graphs.
\newblock In {\em Proceedings of the AAAI Conference on Artificial Intelligence}, volume~37, pages 6718--6727, 2023.

\bibitem[\protect\citeauthoryear{B{\aa}nkestad \bgroup \em et al.\egroup }{2024}]{baankestad2024ising}
Maria B{\aa}nkestad, Jennifer Andersson, Sebastian Mair, and Jens Sj{\"o}lund.
\newblock Ising on the graph: Task-specific graph subsampling via the ising model.
\newblock {\em arXiv preprint arXiv:2402.10206}, 2024.

\bibitem[\protect\citeauthoryear{Baswana and Sen}{2003}]{baswana2003simplespanner}
Surender Baswana and Sandeep Sen.
\newblock A simple linear time algorithm for computing a (2 k—1)-spanner of o (n 1+ 1/k) size in weighted graphs.
\newblock In {\em Automata, Languages and Programming: 30th International Colloquium, ICALP 2003 Eindhoven, The Netherlands, June 30--July 4, 2003 Proceedings 30}, pages 384--396. Springer, 2003.

\bibitem[\protect\citeauthoryear{Batson \bgroup \em et al.\egroup }{2009}]{batson2009twice}
Joshua~D Batson, Daniel~A Spielman, and Nikhil Srivastava.
\newblock Twice-ramanujan sparsifiers.
\newblock In {\em Proceedings of the forty-first annual ACM symposium on Theory of computing}, pages 255--262, 2009.

\bibitem[\protect\citeauthoryear{Batson \bgroup \em et al.\egroup }{2013}]{batson2013sparsifier}
Joshua Batson, Daniel~A Spielman, Nikhil Srivastava, and Shang-Hua Teng.
\newblock Spectral sparsification of graphs: theory and algorithms.
\newblock {\em Communications of the ACM}, 56(8):87--94, 2013.

\bibitem[\protect\citeauthoryear{Beg \bgroup \em et al.\egroup }{2018}]{beg2018scalableapproximation}
Maham~Anwar Beg, Muhammad Ahmad, Arif Zaman, and Imdadullah Khan.
\newblock Scalable approximation algorithm for graph summarization.
\newblock In {\em Advances in Knowledge Discovery and Data Mining: 22nd Pacific-Asia Conference, PAKDD 2018, Melbourne, VIC, Australia, June 3-6, 2018, Proceedings, Part III 22}, pages 502--514. Springer, 2018.

\bibitem[\protect\citeauthoryear{Bing \bgroup \em et al.\egroup }{2023}]{bing2023heterogeneous}
Rui Bing, Guan Yuan, Mu~Zhu, Fanrong Meng, Huifang Ma, and Shaojie Qiao.
\newblock Heterogeneous graph neural networks analysis: a survey of techniques, evaluations and applications.
\newblock {\em Artificial Intelligence Review}, 56(8):8003--8042, 2023.

\bibitem[\protect\citeauthoryear{Bonchi \bgroup \em et al.\egroup }{2014}]{bonchi2014identity}
Francesco Bonchi, Aristides Gionis, and Tamir Tassa.
\newblock Identity obfuscation in graphs through the information theoretic lens.
\newblock {\em Information Sciences}, 275:232--256, 2014.

\bibitem[\protect\citeauthoryear{Brandeis \bgroup \em et al.\egroup }{2020}]{brandeis2020kcorewalk}
Simon Brandeis, Adrian Jarret, and Pierre Sevestre.
\newblock About graph degeneracy, representation learning and scalability.
\newblock {\em arXiv e-prints}, pages arXiv--2009, 2020.

\bibitem[\protect\citeauthoryear{Bravo~Hermsdorff and Gunderson}{2019}]{bravo2019unifyingsparse&coarse}
Gecia Bravo~Hermsdorff and Lee Gunderson.
\newblock A unifying framework for spectrum-preserving graph sparsification and coarsening.
\newblock {\em Advances in Neural Information Processing Systems}, 32, 2019.

\bibitem[\protect\citeauthoryear{Buffelli \bgroup \em et al.\egroup }{2022}]{buffelli2022sizeshiftreg}
Davide Buffelli, Pietro Li{\`o}, and Fabio Vandin.
\newblock Sizeshiftreg: a regularization method for improving size-generalization in graph neural networks.
\newblock {\em Advances in Neural Information Processing Systems}, 35:31871--31885, 2022.

\bibitem[\protect\citeauthoryear{Cai \bgroup \em et al.\egroup }{2021}]{cai2021goren}
Chen Cai, Dingkang Wang, and Yusu Wang.
\newblock Graph coarsening with neural networks.
\newblock In {\em 9th International conference on Learning Representations}, 2021.

\bibitem[\protect\citeauthoryear{Cao \bgroup \em et al.\egroup }{2024}]{cao2024graph}
Linfeng Cao, Haoran Deng, Chunping Wang, Lei Chen, and Yang Yang.
\newblock Graph-skeleton:\~{} 1\% nodes are sufficient to represent billion-scale graph.
\newblock In {\em WWW}, 2024.

\bibitem[\protect\citeauthoryear{Chen and Safro}{2011}]{chen2011algebraicad}
Jie Chen and Ilya Safro.
\newblock Algebraic distance on graphs.
\newblock {\em SIAM Journal on Scientific Computing}, 33(6):3468--3490, 2011.

\bibitem[\protect\citeauthoryear{Chen \bgroup \em et al.\egroup }{2018}]{chen2018harp}
Haochen Chen, Bryan Perozzi, Yifan Hu, and Steven Skiena.
\newblock Harp: Hierarchical representation learning for networks.
\newblock In {\em Proceedings of the AAAI conference on artificial intelligence}, volume~32, 2018.

\bibitem[\protect\citeauthoryear{Chen \bgroup \em et al.\egroup }{2021}]{chen2021sparseloottery}
Tianlong Chen, Yongduo Sui, Xuxi Chen, Aston Zhang, and Zhangyang Wang.
\newblock A unified lottery ticket hypothesis for graph neural networks.
\newblock In {\em International conference on machine learning}, pages 1695--1706. PMLR, 2021.

\bibitem[\protect\citeauthoryear{Chen \bgroup \em et al.\egroup }{2022}]{chen2022graphscsurvey}
Jie Chen, Yousef Saad, and Zechen Zhang.
\newblock Graph coarsening: from scientific computing to machine learning.
\newblock {\em SeMA Journal}, pages 1--37, 2022.

\bibitem[\protect\citeauthoryear{Chung}{2005}]{chung2005laplacianssysmentricdirected}
Fan Chung.
\newblock Laplacians and the cheeger inequality for directed graphs.
\newblock {\em Annals of Combinatorics}, 9:1--19, 2005.

\bibitem[\protect\citeauthoryear{Das \bgroup \em et al.\egroup }{2016}]{das2016distribution}
Kinkar~Ch Das, Seyed~Ahmad Mojallal, and Vilmar Trevisan.
\newblock Distribution of laplacian eigenvalues of graphs.
\newblock {\em Linear Algebra and its Applications}, 508:48--61, 2016.

\bibitem[\protect\citeauthoryear{De~Lange \bgroup \em et al.\egroup }{2021}]{de2021continual}
Matthias De~Lange, Rahaf Aljundi, Marc Masana, Sarah Parisot, Xu~Jia, Ale{\v{s}} Leonardis, Gregory Slabaugh, and Tinne Tuytelaars.
\newblock A continual learning survey: Defying forgetting in classification tasks.
\newblock {\em IEEE transactions on pattern analysis and machine intelligence}, 44(7):3366--3385, 2021.

\bibitem[\protect\citeauthoryear{Dhillon \bgroup \em et al.\egroup }{2007}]{dhillon2007heavy}
Inderjit~S Dhillon, Yuqiang Guan, and Brian Kulis.
\newblock Weighted graph cuts without eigenvectors a multilevel approach.
\newblock {\em IEEE transactions on pattern analysis and machine intelligence}, 29(11):1944--1957, 2007.

\bibitem[\protect\citeauthoryear{Dickens \bgroup \em et al.\egroup }{2023}]{dickens2023convolutionmatch}
Charles Dickens, Eddie Huang, Aishwarya Reganti, Jiong Zhu, Karthik Subbian, and Danai Koutra.
\newblock Graph coarsening via convolution matching for scalable graph neural network training.
\newblock {\em arXiv preprint arXiv:2312.15520}, 2023.

\bibitem[\protect\citeauthoryear{Ding \bgroup \em et al.\egroup }{2022}]{ding2022faster}
Mucong Ding, Xiaoyu Liu, Tahseen Rabbani, Teresa Ranadive, Tai-Ching Tuan, and Furong Huang.
\newblock Faster hyperparameter search for gnns via calibrated dataset condensation.
\newblock 2022.

\bibitem[\protect\citeauthoryear{Dong \bgroup \em et al.\egroup }{2022}]{dong2022privacyforfree}
Tian Dong, Bo~Zhao, and Lingjuan Lyu.
\newblock Privacy for free: How does dataset condensation help privacy?
\newblock In {\em International Conference on Machine Learning}, pages 5378--5396. PMLR, 2022.

\bibitem[\protect\citeauthoryear{Dorfler and Bullo}{2012}]{dorfler2012kron}
Florian Dorfler and Francesco Bullo.
\newblock Kron reduction of graphs with applications to electrical networks.
\newblock {\em IEEE Transactions on Circuits and Systems I: Regular Papers}, 60(1):150--163, 2012.

\bibitem[\protect\citeauthoryear{Du \bgroup \em et al.\egroup }{2019}]{du2019graph}
Simon~S Du, Kangcheng Hou, Russ~R Salakhutdinov, Barnabas Poczos, Ruosong Wang, and Keyulu Xu.
\newblock Graph neural tangent kernel: Fusing graph neural networks with graph kernels.
\newblock {\em Advances in neural information processing systems}, 32, 2019.

\bibitem[\protect\citeauthoryear{Elsken \bgroup \em et al.\egroup }{2019}]{elsken2019neural}
Thomas Elsken, Jan~Hendrik Metzen, and Frank Hutter.
\newblock Neural architecture search: A survey.
\newblock {\em The Journal of Machine Learning Research}, 20(1):1997--2017, 2019.

\bibitem[\protect\citeauthoryear{Fan \bgroup \em et al.\egroup }{2019}]{fan2019socialreco}
Wenqi Fan, Yao Ma, Qing Li, Yuan He, Eric Zhao, Jiliang Tang, and Dawei Yin.
\newblock Graph neural networks for social recommendation.
\newblock In {\em The world wide web conference}, pages 417--426, 2019.

\bibitem[\protect\citeauthoryear{Fang \bgroup \em et al.\egroup }{2010}]{fang2010multilevelindependentedge}
Haw-ren Fang, Sophia Sakellaridi, and Yousef Saad.
\newblock Multilevel manifold learning with application to spectral clustering.
\newblock In {\em Proceedings of the 19th ACM international conference on Information and knowledge management}, pages 419--428, 2010.

\bibitem[\protect\citeauthoryear{Fang \bgroup \em et al.\egroup }{2024}]{fangjf_exgc}
Junfeng Fang, Xinglin Li, Yongduo Sui, Yuan Gao, Guibin Zhang, Kun Wang, Xiang Wang, and Xiangnan He.
\newblock Exgc: Bridging efficiency and explainability in graph condensation.
\newblock In {\em WWW}, 2024.

\bibitem[\protect\citeauthoryear{Feng \bgroup \em et al.\egroup }{2023}]{feng2023fair}
Qizhang Feng, Zhimeng Jiang, Ruiquan Li, Yicheng Wang, Na~Zou, Jiang Bian, and Xia Hu.
\newblock Fair graph distillation.
\newblock In {\em Thirty-seventh Conference on Neural Information Processing Systems}, 2023.

\bibitem[\protect\citeauthoryear{Feng}{2016}]{feng2016spectral}
Zhuo Feng.
\newblock Spectral graph sparsification in nearly-linear time leveraging efficient spectral perturbation analysis.
\newblock In {\em Proceedings of the 53rd Annual Design Automation Conference}, pages 1--6, 2016.

\bibitem[\protect\citeauthoryear{Gao and Wu}{2023}]{gao2023multiple}
Jian Gao and Jianshe Wu.
\newblock Multiple sparse graphs condensation.
\newblock {\em Knowledge-Based Systems}, 278:110904, 2023.

\bibitem[\protect\citeauthoryear{Gao \bgroup \em et al.\egroup }{2023}]{gao2023graph}
Xinyi Gao, Tong Chen, Yilong Zang, Wentao Zhang, Quoc Viet~Hung Nguyen, Kai Zheng, and Hongzhi Yin.
\newblock Graph condensation for inductive node representation learning.
\newblock {\em arXiv preprint arXiv:2307.15967}, 2023.

\bibitem[\protect\citeauthoryear{Garg and Jaakkola}{2019}]{garg2019solving}
Vikas Garg and Tommi Jaakkola.
\newblock Solving graph compression via optimal transport.
\newblock {\em Advances in Neural Information Processing Systems}, 32, 2019.

\bibitem[\protect\citeauthoryear{Generale \bgroup \em et al.\egroup }{2022}]{generale2022scalingrgcn}
Alessandro Generale, Till Blume, and Michael Cochez.
\newblock Scaling r-gcn training with graph summarization.
\newblock In {\em Companion Proceedings of the Web Conference 2022}, pages 1073--1082, 2022.

\bibitem[\protect\citeauthoryear{Geng \bgroup \em et al.\egroup }{2023}]{geng2023surveydatadistillation}
Jiahui Geng, Zongxiong Chen, Yuandou Wang, Herbert Woisetschlaeger, Sonja Schimmler, Ruben Mayer, Zhiming Zhao, and Chunming Rong.
\newblock A survey on dataset distillation: Approaches, applications and future directions.
\newblock {\em arXiv preprint arXiv:2305.01975}, 2023.

\bibitem[\protect\citeauthoryear{Gligorijevi{\'c} \bgroup \em et al.\egroup }{2021}]{gligorijevic2021structure}
Vladimir Gligorijevi{\'c}, P~Douglas Renfrew, Tomasz Kosciolek, Julia~Koehler Leman, Daniel Berenberg, Tommi Vatanen, Chris Chandler, Bryn~C Taylor, Ian~M Fisk, Hera Vlamakis, et~al.
\newblock Structure-based protein function prediction using graph convolutional networks.
\newblock {\em Nature communications}, 12(1):3168, 2021.

\bibitem[\protect\citeauthoryear{Gong \bgroup \em et al.\egroup }{2023}]{gong2023neighborhood}
Shengbo Gong, Jiajun Zhou, Chenxuan Xie, and Qi~Xuan.
\newblock Neighborhood homophily-based graph convolutional network.
\newblock In {\em Proceedings of the 32nd ACM International Conference on Information and Knowledge Management}, CIKM '23, 2023.

\bibitem[\protect\citeauthoryear{Gupta \bgroup \em et al.\egroup }{2023}]{gupta2023mirage}
Mridul Gupta, Sahil Manchanda, Sayan Ranu, and Hariprasad Kodamana.
\newblock Mirage: Model-agnostic graph distillation for graph classification.
\newblock {\em arXiv preprint arXiv:2310.09486}, 2023.

\bibitem[\protect\citeauthoryear{Gutman and Zhou}{2006}]{gutman2006laplacian}
Ivan Gutman and Bo~Zhou.
\newblock Laplacian energy of a graph.
\newblock {\em Linear Algebra and its applications}, 414(1):29--37, 2006.

\bibitem[\protect\citeauthoryear{He \bgroup \em et al.\egroup }{2021}]{he2021distmile}
Yuntian He, Saket Gurukar, Pouya Kousha, Hari Subramoni, Dhabaleswar~K Panda, and Srinivasan Parthasarathy.
\newblock Distmile: a distributed multi-level framework for scalable graph embedding.
\newblock In {\em 2021 IEEE 28th International Conference on High Performance Computing, Data, and Analytics (HiPC)}, pages 282--291. IEEE, 2021.

\bibitem[\protect\citeauthoryear{Hu \bgroup \em et al.\egroup }{2021}]{hu2021ogb}
Weihua Hu, Matthias Fey, Hongyu Ren, Maho Nakata, Yuxiao Dong, and Jure Leskovec.
\newblock Ogb-lsc: A large-scale challenge for machine learning on graphs.
\newblock {\em NeurIPS}, 34, 2021.

\bibitem[\protect\citeauthoryear{Huang \bgroup \em et al.\egroup }{2021}]{huang2021scaling}
Zengfeng Huang, Shengzhong Zhang, Chong Xi, Tang Liu, and Min Zhou.
\newblock Scaling up graph neural networks via graph coarsening.
\newblock In {\em Proceedings of the 27th ACM SIGKDD conference on knowledge discovery \& data mining}, pages 675--684, 2021.

\bibitem[\protect\citeauthoryear{Imre \bgroup \em et al.\egroup }{2020}]{imre2020spectrumvisualize}
Martin Imre, Jun Tao, Yongyu Wang, Zhiqiang Zhao, Zhuo Feng, and Chaoli Wang.
\newblock Spectrum-preserving sparsification for visualization of big graphs.
\newblock {\em Computers \& Graphics}, 87:89--102, 2020.

\bibitem[\protect\citeauthoryear{Interdonato \bgroup \em et al.\egroup }{2020}]{interdonato2020multilayer}
Roberto Interdonato, Matteo Magnani, Diego Perna, Andrea Tagarelli, and Davide Vega.
\newblock Multilayer network simplification: approaches, models and methods.
\newblock {\em Computer Science Review}, 36:100246, 2020.

\bibitem[\protect\citeauthoryear{Itzkovitz \bgroup \em et al.\egroup }{2005}]{itzkovitz2005coarsecomplexnetworks}
Shalev Itzkovitz, Reuven Levitt, Nadav Kashtan, Ron Milo, Michael Itzkovitz, and Uri Alon.
\newblock Coarse-graining and self-dissimilarity of complex networks.
\newblock {\em Physical Review E}, 71(1):016127, 2005.

\bibitem[\protect\citeauthoryear{Jang \bgroup \em et al.\egroup }{2016}]{jang2016repara}
Eric Jang, Shixiang Gu, and Ben Poole.
\newblock Categorical reparameterization with gumbel-softmax.
\newblock In {\em International Conference on Learning Representations}, 2016.

\bibitem[\protect\citeauthoryear{Jia \bgroup \em et al.\egroup }{2020}]{jia2020redundancyfree}
Zhihao Jia, Sina Lin, Rex Ying, Jiaxuan You, Jure Leskovec, and Alex Aiken.
\newblock Redundancy-free computation for graph neural networks.
\newblock In {\em Proceedings of the 26th ACM SIGKDD International Conference on Knowledge Discovery \& Data Mining}, pages 997--1005, 2020.

\bibitem[\protect\citeauthoryear{Jin \bgroup \em et al.\egroup }{2022a}]{jin2022doscondensing}
Wei Jin, Xianfeng Tang, Haoming Jiang, Zheng Li, Danqing Zhang, Jiliang Tang, and Bing Yin.
\newblock Condensing graphs via one-step gradient matching.
\newblock In {\em Proceedings of the 28th ACM SIGKDD Conference on Knowledge Discovery and Data Mining}, pages 720--730, 2022.

\bibitem[\protect\citeauthoryear{Jin \bgroup \em et al.\egroup }{2022b}]{jin2021gcond}
Wei Jin, Lingxiao Zhao, Shichang Zhang, Yozen Liu, Jiliang Tang, and Neil Shah.
\newblock Graph condensation for graph neural networks.
\newblock In {\em International Conference on Learning Representations}, 2022.

\bibitem[\protect\citeauthoryear{Karypis and Kumar}{1997}]{karypis1997metis}
George Karypis and Vipin Kumar.
\newblock Metis: A software package for partitioning unstructured graphs, partitioning meshes, and computing fill-reducing orderings of sparse matrices.
\newblock 1997.

\bibitem[\protect\citeauthoryear{Kemker and Kanan}{2018}]{kemker2018replay}
Ronald Kemker and Christopher Kanan.
\newblock Fearnet: Brain-inspired model for incremental learning.
\newblock In {\em International Conference on Learning Representations}, 2018.

\bibitem[\protect\citeauthoryear{Khan \bgroup \em et al.\egroup }{2015}]{khan2015lossless2}
Kifayat~Ullah Khan, Waqas Nawaz, and Young-Koo Lee.
\newblock Set-based approximate approach for lossless graph summarization.
\newblock {\em Computing}, 97:1185--1207, 2015.

\bibitem[\protect\citeauthoryear{Kipf and Welling}{2016}]{kipf2016semi}
Thomas~N Kipf and Max Welling.
\newblock Semi-supervised classification with graph convolutional networks.
\newblock In {\em International Conference on Learning Representations}, 2016.

\bibitem[\protect\citeauthoryear{Kumar \bgroup \em et al.\egroup }{2023}]{kumar2023fgc1}
Manoj Kumar, Anurag Sharma, Shashwat Saxena, and Sandeep Kumar.
\newblock Featured graph coarsening with similarity guarantees.
\newblock In {\em International Conference on Machine Learning}, pages 17953--17975. PMLR, 2023.

\bibitem[\protect\citeauthoryear{Langville and Meyer}{2004}]{langville2004deeperpagerank}
Amy~N Langville and Carl~D Meyer.
\newblock Deeper inside pagerank.
\newblock {\em Internet Mathematics}, 1(3), 2004.

\bibitem[\protect\citeauthoryear{Lee and Sun}{2018}]{lee2018twicelinear}
Yin~Tat Lee and He~Sun.
\newblock Constructing linear-sized spectral sparsification in almost-linear time.
\newblock {\em SIAM Journal on Computing}, 47(6):2315--2336, 2018.

\bibitem[\protect\citeauthoryear{LeFevre and Terzi}{2010}]{lefevre2010grass1}
Kristen LeFevre and Evimaria Terzi.
\newblock Grass: Graph structure summarization.
\newblock In {\em Proceedings of the 2010 SIAM International Conference on Data Mining}, pages 454--465. SIAM, 2010.

\bibitem[\protect\citeauthoryear{Li \bgroup \em et al.\egroup }{2022}]{li2022out}
Haoyang Li, Xin Wang, Ziwei Zhang, and Wenwu Zhu.
\newblock Out-of-distribution generalization on graphs: A survey.
\newblock {\em arXiv preprint arXiv:2202.07987}, 2022.

\bibitem[\protect\citeauthoryear{Li \bgroup \em et al.\egroup }{2023a}]{braingnn-igs}
Gaotang Li, Marlena Duda, Xiang Zhang, Danai Koutra, and Yujun Yan.
\newblock Interpretable sparsification of brain graphs: Better practices and effective designs for graph neural networks.
\newblock In {\em Proceedings of the 29th ACM SIGKDD Conference on Knowledge Discovery and Data Mining}, KDD '23, 2023.

\bibitem[\protect\citeauthoryear{Li \bgroup \em et al.\egroup }{2023b}]{li2023attend}
Xinglin Li, Kun Wang, Hanhui Deng, Yuxuan Liang, and Di~Wu.
\newblock Attend who is weak: Enhancing graph condensation via cross-free adversarial training.
\newblock {\em arXiv preprint arXiv:2311.15772}, 2023.

\bibitem[\protect\citeauthoryear{Liang \bgroup \em et al.\egroup }{2021}]{liang2021mile}
Jiongqian Liang, Saket Gurukar, and Srinivasan Parthasarathy.
\newblock Mile: A multi-level framework for scalable graph embedding.
\newblock In {\em Proceedings of the International AAAI Conference on Web and Social Media}, volume~15, pages 361--372, 2021.

\bibitem[\protect\citeauthoryear{Liu \bgroup \em et al.\egroup }{2018}]{liu2018summarization}
Yike Liu, Tara Safavi, Abhilash Dighe, and Danai Koutra.
\newblock Graph summarization methods and applications: A survey.
\newblock {\em ACM computing surveys (CSUR)}, 51(3):1--34, 2018.

\bibitem[\protect\citeauthoryear{Liu \bgroup \em et al.\egroup }{2021}]{liu2021exact}
Zirui Liu, Kaixiong Zhou, Fan Yang, Li~Li, Rui Chen, and Xia Hu.
\newblock Exact: Scalable graph neural networks training via extreme activation compression.
\newblock In {\em International Conference on Learning Representations}, 2021.

\bibitem[\protect\citeauthoryear{Liu \bgroup \em et al.\egroup }{2022}]{liu2022graph}
Mengyang Liu, Shanchuan Li, Xinshi Chen, and Le~Song.
\newblock Graph condensation via receptive field distribution matching.
\newblock {\em arXiv preprint arXiv:2206.13697}, 2022.

\bibitem[\protect\citeauthoryear{Liu \bgroup \em et al.\egroup }{2023a}]{liu2023cgp}
Chuang Liu, Xueqi Ma, Yibing Zhan, Liang Ding, Dapeng Tao, Bo~Du, Wenbin Hu, and Danilo~P Mandic.
\newblock Comprehensive graph gradual pruning for sparse training in graph neural networks.
\newblock {\em IEEE Transactions on Neural Networks and Learning Systems}, 2023.

\bibitem[\protect\citeauthoryear{Liu \bgroup \em et al.\egroup }{2023b}]{liu2023graph}
Yang Liu, Deyu Bo, and Chuan Shi.
\newblock Graph condensation via eigenbasis matching.
\newblock {\em arXiv preprint arXiv:2310.09202}, 2023.

\bibitem[\protect\citeauthoryear{Liu \bgroup \em et al.\egroup }{2023c}]{liu2023cat}
Yilun Liu, Ruihong Qiu, and Zi~Huang.
\newblock Cat: Balanced continual graph learning with graph condensation.
\newblock {\em arXiv preprint arXiv:2309.09455}, 2023.

\bibitem[\protect\citeauthoryear{Liu \bgroup \em et al.\egroup }{2023d}]{liu2023puma}
Yilun Liu, Ruihong Qiu, Yanran Tang, Hongzhi Yin, and Zi~Huang.
\newblock Puma: Efficient continual graph learning with graph condensation.
\newblock {\em arXiv preprint arXiv:2312.14439}, 2023.

\bibitem[\protect\citeauthoryear{Liu \bgroup \em et al.\egroup }{2024}]{liu2024review}
Zewen Liu, Guancheng Wan, B~Aditya Prakash, Max~SY Lau, and Wei Jin.
\newblock A review of graph neural networks in epidemic modeling.
\newblock 2024.

\bibitem[\protect\citeauthoryear{Livne and Brandt}{2012}]{livne2012affinity}
Oren~E Livne and Achi Brandt.
\newblock Lean algebraic multigrid (lamg): Fast graph laplacian linear solver.
\newblock {\em SIAM Journal on Scientific Computing}, 34(4):B499--B522, 2012.

\bibitem[\protect\citeauthoryear{Loukas and Vandergheynst}{2018}]{loukas2018spectrally}
Andreas Loukas and Pierre Vandergheynst.
\newblock Spectrally approximating large graphs with smaller graphs.
\newblock In {\em International Conference on Machine Learning}, pages 3237--3246. PMLR, 2018.

\bibitem[\protect\citeauthoryear{Loukas}{2019}]{loukas2019spectral&cut}
Andreas Loukas.
\newblock Graph reduction with spectral and cut guarantees.
\newblock {\em J. Mach. Learn. Res.}, 20(116):1--42, 2019.

\bibitem[\protect\citeauthoryear{Ma and Tang}{2021}]{ma2021deep}
Yao Ma and Jiliang Tang.
\newblock {\em Deep learning on graphs}.
\newblock Cambridge University Press, 2021.

\bibitem[\protect\citeauthoryear{Mao \bgroup \em et al.\egroup }{2023}]{mao2023gcare}
Runze Mao, Wenqi Fan, and Qing Li.
\newblock Gcare: Mitigating subgroup unfairness in graph condensation through adversarial regularization.
\newblock {\em Applied Sciences}, 13(16):9166, 2023.

\bibitem[\protect\citeauthoryear{Mehrabi \bgroup \em et al.\egroup }{2021}]{mehrabi2021survey}
Ninareh Mehrabi, Fred Morstatter, Nripsuta Saxena, Kristina Lerman, and Aram Galstyan.
\newblock A survey on bias and fairness in machine learning.
\newblock {\em ACM computing surveys (CSUR)}, 54(6):1--35, 2021.

\bibitem[\protect\citeauthoryear{Navlakha \bgroup \em et al.\egroup }{2008}]{navlakha2008lossless1}
Saket Navlakha, Rajeev Rastogi, and Nisheeth Shrivastava.
\newblock Graph summarization with bounded error.
\newblock In {\em Proceedings of the 2008 ACM SIGMOD international conference on Management of data}, pages 419--432, 2008.

\bibitem[\protect\citeauthoryear{Nguyen \bgroup \em et al.\egroup }{2020}]{nguyen2020dataset}
Timothy Nguyen, Zhourong Chen, and Jaehoon Lee.
\newblock Dataset meta-learning from kernel ridge-regression.
\newblock {\em arXiv preprint arXiv:2011.00050}, 2020.

\bibitem[\protect\citeauthoryear{Nguyen \bgroup \em et al.\egroup }{2021}]{nguyen2021dataset}
Timothy Nguyen, Roman Novak, Lechao Xiao, and Jaehoon Lee.
\newblock Dataset distillation with infinitely wide convolutional networks.
\newblock {\em Advances in Neural Information Processing Systems}, 34:5186--5198, 2021.

\bibitem[\protect\citeauthoryear{Nian \bgroup \em et al.\egroup }{2023}]{nian2023GDMexplain}
Yi~Nian, Wei Jin, and Lu~Lin.
\newblock In-process global interpretation for graph learning via distribution matching.
\newblock {\em arXiv preprint arXiv:2306.10447}, 2023.

\bibitem[\protect\citeauthoryear{Pan \bgroup \em et al.\egroup }{2023}]{pan2023fedgkd}
Qiying Pan, Ruofan Wu, Tengfei Liu, Tianyi Zhang, Yifei Zhu, and Weiqiang Wang.
\newblock Fedgkd: Unleashing the power of collaboration in federated graph neural networks.
\newblock {\em arXiv preprint arXiv:2309.09517}, 2023.

\bibitem[\protect\citeauthoryear{Peyr{\'e} \bgroup \em et al.\egroup }{2019}]{peyre2019computational}
Gabriel Peyr{\'e}, Marco Cuturi, et~al.
\newblock Computational optimal transport: With applications to data science.
\newblock {\em Foundations and Trends{\textregistered} in Machine Learning}, 11(5-6):355--607, 2019.

\bibitem[\protect\citeauthoryear{Purohit \bgroup \em et al.\egroup }{2014}]{purohit2014coarsenet}
Manish Purohit, B~Aditya Prakash, Chanhyun Kang, Yao Zhang, and VS~Subrahmanian.
\newblock Fast influence-based coarsening for large networks.
\newblock In {\em Proceedings of the 20th ACM SIGKDD international conference on Knowledge discovery and data mining}, pages 1296--1305, 2014.

\bibitem[\protect\citeauthoryear{Razin \bgroup \em et al.\egroup }{2023}]{razin2023WIS}
Noam Razin, Tom Verbin, and Nadav Cohen.
\newblock On the ability of graph neural networks to model interactions between vertices.
\newblock In {\em Advances in Neural Information Processing Systems}, 2023.

\bibitem[\protect\citeauthoryear{Reiser \bgroup \em et al.\egroup }{2022}]{reiser2022graph}
Patrick Reiser, Marlen Neubert, Andr{\'e} Eberhard, Luca Torresi, Chen Zhou, Chen Shao, Houssam Metni, Clint van Hoesel, Henrik Schopmans, Timo Sommer, et~al.
\newblock Graph neural networks for materials science and chemistry.
\newblock {\em Communications Materials}, 3(1):93, 2022.

\bibitem[\protect\citeauthoryear{Ren \bgroup \em et al.\egroup }{2021}]{ren2021comprehensive}
Pengzhen Ren, Yun Xiao, Xiaojun Chang, Po-Yao Huang, Zhihui Li, Xiaojiang Chen, and Xin Wang.
\newblock A comprehensive survey of neural architecture search: Challenges and solutions.
\newblock {\em ACM Computing Surveys (CSUR)}, 54(4):1--34, 2021.

\bibitem[\protect\citeauthoryear{Riondato \bgroup \em et al.\egroup }{2017}]{riondato2017S2l}
Matteo Riondato, David Garc{\'\i}a-Soriano, and Francesco Bonchi.
\newblock Graph summarization with quality guarantees.
\newblock {\em Data mining and knowledge discovery}, 31:314--349, 2017.

\bibitem[\protect\citeauthoryear{Rong \bgroup \em et al.\egroup }{2019}]{rong2019dropedge}
Yu~Rong, Wenbing Huang, Tingyang Xu, and Junzhou Huang.
\newblock Dropedge: Towards deep graph convolutional networks on node classification.
\newblock {\em arXiv preprint arXiv:1907.10903}, 2019.

\bibitem[\protect\citeauthoryear{Sachdeva and McAuley}{2023}]{sachdeva2023DDasurvey}
Noveen Sachdeva and Julian McAuley.
\newblock Data distillation: A survey.
\newblock {\em arXiv preprint arXiv:2301.04272}, 2023.

\bibitem[\protect\citeauthoryear{Safro \bgroup \em et al.\egroup }{2015}]{safro2015advancedcoarsenpartition}
Ilya Safro, Peter Sanders, and Christian Schulz.
\newblock Advanced coarsening schemes for graph partitioning.
\newblock {\em Journal of Experimental Algorithmics (JEA)}, 19:1--24, 2015.

\bibitem[\protect\citeauthoryear{Salha \bgroup \em et al.\egroup }{2019}]{salha2019kcoredegeneracy}
Guillaume Salha, Romain Hennequin, Viet~Anh Tran, and Michalis Vazirgiannis.
\newblock A degeneracy framework for scalable graph autoencoders.
\newblock In {\em Proceedings of the 28th International Joint Conference on Artificial Intelligence}, pages 3353--3359, 2019.

\bibitem[\protect\citeauthoryear{Schlichtkrull \bgroup \em et al.\egroup }{2018}]{schlichtkrull2018rgcn}
Michael Schlichtkrull, Thomas~N Kipf, Peter Bloem, Rianne Van Den~Berg, Ivan Titov, and Max Welling.
\newblock Modeling relational data with graph convolutional networks.
\newblock In {\em The Semantic Web: 15th International Conference, ESWC 2018, Heraklion, Crete, Greece, June 3--7, 2018, Proceedings 15}, pages 593--607. Springer, 2018.

\bibitem[\protect\citeauthoryear{Sener and Savarese}{2018}]{sener2018activecoreset}
Ozan Sener and Silvio Savarese.
\newblock Active learning for convolutional neural networks: A core-set approach.
\newblock In {\em International Conference on Learning Representations}, 2018.

\bibitem[\protect\citeauthoryear{Seo \bgroup \em et al.\egroup }{2024}]{seo2024teddy}
Hyunjin Seo, Jihun Yun, and Eunho Yang.
\newblock {TEDDY}: Trimming edges with degree-based graph diffusion strategy.
\newblock In {\em The Twelfth International Conference on Learning Representations}, 2024.

\bibitem[\protect\citeauthoryear{Shabani \bgroup \em et al.\egroup }{2023}]{shabani2023surveyGNN}
Nasrin Shabani, Jia Wu, Amin Beheshti, Jin Foo, Ambreen Hanif, and Maryam Shahabikargar.
\newblock A survey on graph neural networks for graph summarization.
\newblock {\em arXiv preprint arXiv:2302.06114}, 2023.

\bibitem[\protect\citeauthoryear{Shi and Weninger}{2017}]{shi2017KG}
Baoxu Shi and Tim Weninger.
\newblock Proje: Embedding projection for knowledge graph completion.
\newblock In {\em Proceedings of the AAAI Conference on Artificial Intelligence}, volume~31, 2017.

\bibitem[\protect\citeauthoryear{Shokri \bgroup \em et al.\egroup }{2017}]{shokri2017membership}
Reza Shokri, Marco Stronati, Congzheng Song, and Vitaly Shmatikov.
\newblock Membership inference attacks against machine learning models.
\newblock In {\em 2017 IEEE symposium on security and privacy (SP)}, pages 3--18. IEEE, 2017.

\bibitem[\protect\citeauthoryear{Shuman \bgroup \em et al.\egroup }{2015}]{shuman2015pyramid}
David~I Shuman, Mohammad~Javad Faraji, and Pierre Vandergheynst.
\newblock A multiscale pyramid transform for graph signals.
\newblock {\em IEEE Transactions on Signal Processing}, 64(8):2119--2134, 2015.

\bibitem[\protect\citeauthoryear{Si \bgroup \em et al.\egroup }{2023}]{si2022serving}
Si~Si, Felix Yu, Ankit~Singh Rawat, Cho-Jui Hsieh, and Sanjiv Kumar.
\newblock Serving graph compression for graph neural networks.
\newblock In {\em International Conference on Learning Representations}, 2023.

\bibitem[\protect\citeauthoryear{Skarding \bgroup \em et al.\egroup }{2021}]{skarding2021dynamicsurvey}
Joakim Skarding, Bogdan Gabrys, and Katarzyna Musial.
\newblock Foundations and modeling of dynamic networks using dynamic graph neural networks: A survey.
\newblock {\em IEEE Access}, 9:79143--79168, 2021.

\bibitem[\protect\citeauthoryear{Song \bgroup \em et al.\egroup }{2023}]{song2023xgcn}
Xiran Song, Jianxun Lian, Hong Huang, Zihan Luo, Wei Zhou, Xue Lin, Mingqi Wu, Chaozhuo Li, Xing Xie, and Hai Jin.
\newblock xgcn: An extreme graph convolutional network for large-scale social link prediction.
\newblock In {\em Proceedings of the ACM Web Conference 2023}, pages 349--359, 2023.

\bibitem[\protect\citeauthoryear{Spielman and Srivastava}{2008}]{spielman2008effectiveresistance}
Daniel~A Spielman and Nikhil Srivastava.
\newblock Graph sparsification by effective resistances.
\newblock In {\em Proceedings of the fortieth annual ACM symposium on Theory of computing}, pages 563--568, 2008.

\bibitem[\protect\citeauthoryear{Such \bgroup \em et al.\egroup }{2017}]{such2017coarsepool}
Felipe~Petroski Such, Shagan Sah, Miguel~Alexander Dominguez, Suhas Pillai, Chao Zhang, Andrew Michael, Nathan~D Cahill, and Raymond Ptucha.
\newblock Robust spatial filtering with graph convolutional neural networks.
\newblock {\em IEEE Journal of Selected Topics in Signal Processing}, 11(6):884--896, 2017.

\bibitem[\protect\citeauthoryear{Sugiyama and Sato}{2023}]{sugiyama2023krondirected}
Tomohiro Sugiyama and Kazuhiro Sato.
\newblock Kron reduction and effective resistance of directed graphs.
\newblock {\em SIAM Journal on Matrix Analysis and Applications}, 44(1):270--292, 2023.

\bibitem[\protect\citeauthoryear{Tian \bgroup \em et al.\egroup }{2008}]{tian2008ksnap}
Yuanyuan Tian, Richard~A Hankins, and Jignesh~M Patel.
\newblock Efficient aggregation for graph summarization.
\newblock In {\em Proceedings of the 2008 ACM SIGMOD international conference on Management of data}, pages 567--580, 2008.

\bibitem[\protect\citeauthoryear{Tsilivis \bgroup \em et al.\egroup }{2022}]{tsilivis2022can}
Nikolaos Tsilivis, Jingtong Su, and Julia Kempe.
\newblock Can we achieve robustness from data alone?
\newblock {\em arXiv preprint arXiv:2207.11727}, 2022.

\bibitem[\protect\citeauthoryear{Tsitsulin \bgroup \em et al.\egroup }{2023}]{tsitsulin2023graphclustegnn}
Anton Tsitsulin, John Palowitch, Bryan Perozzi, and Emmanuel M{\"u}ller.
\newblock Graph clustering with graph neural networks.
\newblock {\em Journal of Machine Learning Research}, 24(127):1--21, 2023.

\bibitem[\protect\citeauthoryear{Wang \bgroup \em et al.\egroup }{2018}]{wang2018dataset}
Tongzhou Wang, Jun-Yan Zhu, Antonio Torralba, and Alexei~A Efros.
\newblock Dataset distillation.
\newblock {\em arXiv preprint arXiv:1811.10959}, 2018.

\bibitem[\protect\citeauthoryear{Wang \bgroup \em et al.\egroup }{2021}]{wang2021leverage}
Hao Wang, Jiaxin Yang, and Jianrong Wang.
\newblock Leverage large-scale biological networks to decipher the genetic basis of human diseases using machine learning.
\newblock {\em Artificial Neural Networks}, pages 229--248, 2021.

\bibitem[\protect\citeauthoryear{Wang \bgroup \em et al.\egroup }{2024}]{wang2023fast}
Lin Wang, Wenqi Fan, Jiatong Li, Yao Ma, and Qing Li.
\newblock Fast graph condensation with structure-based neural tangent kernel.
\newblock {\em Proceedings of the ACM Web Conference}, 2024.

\bibitem[\protect\citeauthoryear{Welling}{2009}]{welling2009herding}
Max Welling.
\newblock Herding dynamical weights to learn.
\newblock In {\em Proceedings of the 26th Annual International Conference on Machine Learning}, pages 1121--1128, 2009.

\bibitem[\protect\citeauthoryear{Wickman \bgroup \em et al.\egroup }{2022}]{wickman2022sparsebyRL}
Ryan Wickman, Xiaofei Zhang, and Weizi Li.
\newblock A generic graph sparsification framework using deep reinforcement learning.
\newblock In {\em 2022 IEEE International Conference on Data Mining (ICDM)}, pages 1221--1226. IEEE, 2022.

\bibitem[\protect\citeauthoryear{Wu and Chen}{2020}]{wu2020graphgan}
Hang-Yang Wu and Yi-Ling Chen.
\newblock Graph sparsification with generative adversarial network.
\newblock In {\em 2020 IEEE International Conference on Data Mining (ICDM)}, pages 1328--1333. IEEE, 2020.

\bibitem[\protect\citeauthoryear{Wu \bgroup \em et al.\egroup }{2014}]{wu2014ag}
Ye~Wu, Zhinong Zhong, Wei Xiong, and Ning Jing.
\newblock Graph summarization for attributed graphs.
\newblock In {\em 2014 International conference on information science, electronics and electrical engineering}, volume~1, pages 503--507. IEEE, 2014.

\bibitem[\protect\citeauthoryear{Wu \bgroup \em et al.\egroup }{2018}]{wu2018leveraging}
Mengmeng Wu, Wanwen Zeng, Wenqiang Liu, Hairong Lv, Ting Chen, and Rui Jiang.
\newblock Leveraging multiple gene networks to prioritize gwas candidate genes via network representation learning.
\newblock {\em Methods}, 145:41--50, 2018.

\bibitem[\protect\citeauthoryear{Wu \bgroup \em et al.\egroup }{2020}]{wu2020comprehensive}
Zonghan Wu, Shirui Pan, Fengwen Chen, Guodong Long, Chengqi Zhang, and S~Yu Philip.
\newblock A comprehensive survey on graph neural networks.
\newblock {\em IEEE transactions on neural networks and learning systems}, 32(1):4--24, 2020.

\bibitem[\protect\citeauthoryear{Wu \bgroup \em et al.\egroup }{2022a}]{wu2022handling}
Qitian Wu, Hengrui Zhang, Junchi Yan, and David Wipf.
\newblock Handling distribution shifts on graphs: An invariance perspective.
\newblock {\em arXiv preprint arXiv:2202.02466}, 2022.

\bibitem[\protect\citeauthoryear{Wu \bgroup \em et al.\egroup }{2022b}]{wu2022reco}
Shiwen Wu, Fei Sun, Wentao Zhang, Xu~Xie, and Bin Cui.
\newblock Graph neural networks in recommender systems: a survey.
\newblock {\em ACM Computing Surveys}, 55(5):1--37, 2022.

\bibitem[\protect\citeauthoryear{Xiao \bgroup \em et al.\egroup }{2024}]{xiao2024disentangled}
Zhenbang Xiao, Shunyu Liu, Yu~Wang, Tongya Zheng, and Mingli Song.
\newblock Disentangled condensation for large-scale graphs.
\newblock {\em arXiv preprint arXiv:2401.12231}, 2024.

\bibitem[\protect\citeauthoryear{Xu \bgroup \em et al.\egroup }{2023}]{xu2023kernel}
Zhe Xu, Yuzhong Chen, Menghai Pan, Huiyuan Chen, Mahashweta Das, Hao Yang, and Hanghang Tong.
\newblock Kernel ridge regression-based graph dataset distillation.
\newblock In {\em Proceedings of the 29th ACM SIGKDD Conference on Knowledge Discovery and Data Mining}, pages 2850--2861, 2023.

\bibitem[\protect\citeauthoryear{Yang \bgroup \em et al.\egroup }{2023}]{yang2023does}
Beining Yang, Kai Wang, Qingyun Sun, Cheng Ji, Xingcheng Fu, Hao Tang, Yang You, and Jianxin Li.
\newblock Does graph distillation see like vision dataset counterpart?
\newblock {\em arXiv preprint arXiv:2310.09192}, 2023.

\bibitem[\protect\citeauthoryear{Ying \bgroup \em et al.\egroup }{2019}]{ying2019gnnexplainer}
Zhitao Ying, Dylan Bourgeois, Jiaxuan You, Marinka Zitnik, and Jure Leskovec.
\newblock Gnnexplainer: Generating explanations for graph neural networks.
\newblock {\em Advances in neural information processing systems}, 32, 2019.

\bibitem[\protect\citeauthoryear{Yuan \bgroup \em et al.\egroup }{2020}]{yuan2020xgnn}
Hao Yuan, Jiliang Tang, Xia Hu, and Shuiwang Ji.
\newblock Xgnn: Towards model-level explanations of graph neural networks.
\newblock In {\em Proceedings of the 26th ACM SIGKDD International Conference on Knowledge Discovery \& Data Mining}, pages 430--438, 2020.

\bibitem[\protect\citeauthoryear{Zeng \bgroup \em et al.\egroup }{2021}]{zeng2021decoupling}
Hanqing Zeng, Muhan Zhang, Yinglong Xia, Ajitesh Srivastava, Andrey Malevich, Rajgopal Kannan, Viktor Prasanna, Long Jin, and Ren Chen.
\newblock Decoupling the depth and scope of graph neural networks.
\newblock {\em Advances in Neural Information Processing Systems}, 34:19665--19679, 2021.

\bibitem[\protect\citeauthoryear{Zha \bgroup \em et al.\egroup }{2023}]{zha2023datacentric}
Daochen Zha, Zaid~Pervaiz Bhat, Kwei-Herng Lai, Fan Yang, and Xia Hu.
\newblock Data-centric ai: Perspectives and challenges.
\newblock In {\em Proceedings of the 2023 SIAM International Conference on Data Mining (SDM)}, pages 945--948. SIAM, 2023.

\bibitem[\protect\citeauthoryear{Zhang \bgroup \em et al.\egroup }{2010}]{zhang2010discoverycancel}
Ning Zhang, Yuanyuan Tian, and Jignesh~M Patel.
\newblock Discovery-driven graph summarization.
\newblock In {\em 2010 IEEE 26th international conference on data engineering (ICDE 2010)}, pages 880--891. IEEE, 2010.

\bibitem[\protect\citeauthoryear{Zhang \bgroup \em et al.\egroup }{2023a}]{zhang2023surveyacceleration}
Shichang Zhang, Atefeh Sohrabizadeh, Cheng Wan, Zijie Huang, Ziniu Hu, Yewen Wang, Jason Cong, Yizhou Sun, et~al.
\newblock A survey on graph neural network acceleration: Algorithms, systems, and customized hardware.
\newblock {\em arXiv preprint arXiv:2306.14052}, 2023.

\bibitem[\protect\citeauthoryear{Zhang \bgroup \em et al.\egroup }{2023b}]{zhang2023ricci}
Xikun Zhang, Dongjin Song, and Dacheng Tao.
\newblock Ricci curvature-based graph sparsification for continual graph representation learning.
\newblock {\em IEEE Transactions on Neural Networks and Learning Systems}, 2023.

\bibitem[\protect\citeauthoryear{Zhang \bgroup \em et al.\egroup }{2024a}]{zhang2024two}
Tianle Zhang, Yuchen Zhang, Kun Wang, Kai Wang, Beining Yang, Kaipeng Zhang, Wenqi Shao, Ping Liu, Joey~Tianyi Zhou, and Yang You.
\newblock Two trades is not baffled: Condense graph via crafting rational gradient matching.
\newblock {\em arXiv preprint arXiv:2402.04924}, 2024.

\bibitem[\protect\citeauthoryear{Zhang \bgroup \em et al.\egroup }{2024b}]{zhang2024Navigating}
Yuchen Zhang, Tianle Zhang, Kai Wang, Ziyao Guo, Yuxuan Liang, Xavier Bresson, Wei Jin, and Yang You.
\newblock Navigating complexity: Toward lossless graph condensation via expanding window matching.
\newblock 2024.

\bibitem[\protect\citeauthoryear{Zhao and Bilen}{2023}]{zhao2023dataset}
Bo~Zhao and Hakan Bilen.
\newblock Dataset condensation with distribution matching.
\newblock In {\em Proceedings of the IEEE/CVF Winter Conference on Applications of Computer Vision}, pages 6514--6523, 2023.

\bibitem[\protect\citeauthoryear{Zhao \bgroup \em et al.\egroup }{2018}]{zhao2018nearlylinearvisualization}
Zhiqiang Zhao, Yongyu Wang, and Zhuo Feng.
\newblock Nearly-linear time spectral graph reduction for scalable graph partitioning and data visualization.
\newblock {\em arXiv preprint arXiv:1812.08942}, 2018.

\bibitem[\protect\citeauthoryear{Zhao \bgroup \em et al.\egroup }{2020}]{zhao2020dataset}
Bo~Zhao, Konda~Reddy Mopuri, and Hakan Bilen.
\newblock Dataset condensation with gradient matching.
\newblock {\em arXiv preprint arXiv:2006.05929}, 2020.

\bibitem[\protect\citeauthoryear{Zhao \bgroup \em et al.\egroup }{2021}]{zhao2021data}
Tong Zhao, Yozen Liu, Leonardo Neves, Oliver Woodford, Meng Jiang, and Neil Shah.
\newblock Data augmentation for graph neural networks.
\newblock In {\em Proceedings of the aaai conference on artificial intelligence}, volume~35, pages 11015--11023, 2021.

\bibitem[\protect\citeauthoryear{Zhao \bgroup \em et al.\egroup }{2022}]{zhao2022graph}
Tong Zhao, Wei Jin, Yozen Liu, Yingheng Wang, Gang Liu, Stephan Günneman, Neil Shah, and Meng Jiang.
\newblock Graph data augmentation for graph machine learning: A survey.
\newblock {\em arXiv preprint arXiv:2202.08871}, 2022.

\bibitem[\protect\citeauthoryear{Zheng \bgroup \em et al.\egroup }{2020}]{zheng2020neuralsparse}
Cheng Zheng, Bo~Zong, Wei Cheng, Dongjin Song, Jingchao Ni, Wenchao Yu, Haifeng Chen, and Wei Wang.
\newblock Robust graph representation learning via neural sparsification.
\newblock In {\em International Conference on Machine Learning}, pages 11458--11468. PMLR, 2020.

\bibitem[\protect\citeauthoryear{Zheng \bgroup \em et al.\egroup }{2022}]{zheng2022graphheter}
Xin Zheng, Yixin Liu, Shirui Pan, Miao Zhang, Di~Jin, and Philip~S Yu.
\newblock Graph neural networks for graphs with heterophily: A survey.
\newblock {\em arXiv preprint arXiv:2202.07082}, 2022.

\bibitem[\protect\citeauthoryear{Zheng \bgroup \em et al.\egroup }{2023a}]{zheng2023towardsdatacentric}
Xin Zheng, Yixin Liu, Zhifeng Bao, Meng Fang, Xia Hu, Alan Wee-Chung Liew, and Shirui Pan.
\newblock Towards data-centric graph machine learning: Review and outlook.
\newblock {\em arXiv preprint arXiv:2309.10979}, 2023.

\bibitem[\protect\citeauthoryear{Zheng \bgroup \em et al.\egroup }{2023b}]{zheng2023structure}
Xin Zheng, Miao Zhang, Chunyang Chen, Quoc Viet~Hung Nguyen, Xingquan Zhu, and Shirui Pan.
\newblock Structure-free graph condensation: From large-scale graphs to condensed graph-free data.
\newblock {\em arXiv preprint arXiv:2306.02664}, 2023.

\bibitem[\protect\citeauthoryear{Zhou \bgroup \em et al.\egroup }{2022}]{zhou2022dataaug}
Jiajun Zhou, Chenxuan Xie, Zhenyu Wen, Xiangyu Zhao, and Qi~Xuan.
\newblock Data augmentation on graphs: A survey.
\newblock {\em arXiv preprint arXiv:2212.09970}, 2022.

\bibitem[\protect\citeauthoryear{Zhu \bgroup \em et al.\egroup }{2020}]{zhu2020beyond}
Jiong Zhu, Yujun Yan, Lingxiao Zhao, Mark Heimann, Leman Akoglu, and Danai Koutra.
\newblock Beyond homophily in graph neural networks: Current limitations and effective designs.
\newblock {\em Advances in neural information processing systems}, 33:7793--7804, 2020.

\bibitem[\protect\citeauthoryear{Zhu \bgroup \em et al.\egroup }{2021}]{zhu2021surveygsl}
Yanqiao Zhu, Weizhi Xu, Jinghao Zhang, Yuanqi Du, Jieyu Zhang, Qiang Liu, Carl Yang, and Shu Wu.
\newblock A survey on graph structure learning: Progress and opportunities.
\newblock {\em arXiv preprint arXiv:2103.03036}, 2021.

\end{thebibliography}

\end{document}